\documentclass[onecolumn,11pt,oneside,draftclsnofoot]{IEEEtran}%

\usepackage{amsbsy}
\usepackage{floatflt} 

\usepackage{amsmath}
\usepackage{amssymb}
\usepackage{times}
\usepackage{graphicx}
\usepackage{xspace}
\usepackage{paralist} 
\usepackage{setspace} 
\usepackage{xypic}
\xyoption{curve}
\usepackage{latexsym}
\usepackage{theorem}
\usepackage{ifthen}
\usepackage{subfigure}
\usepackage{booktabs}

\newcommand{\ceil}[1]{\lceil #1 \rceil }

{\theoremheaderfont{\it} \theorembodyfont{\rmfamily}
\newtheorem{theorem}{Theorem}
\newtheorem{lemma}[theorem]{Lemma}

\newtheorem{corollary}[theorem]{Corollary}

}

\def\eqd{\,{\buildrel d \over =}\,}

\def\ln{{\rm ln}}

\title{Distributed Consensus Algorithms in Sensor Networks With Imperfect
Communication: Link Failures and Channel Noise}
\author{Soummya Kar and Jos\'e M.~F.~Moura$^{*}$
\thanks{The authors are with the Department of Electrical and Computer Engineering,
Carnegie Mellon University, Pittsburgh, PA, USA 15213 (e-mail:
soummyak@andrew.cmu.edu, moura@ece.cmu.edu, ph: (412)268-6341,
fax: (412)268-3890.)}
\thanks{Work supported by the DARPA DSO Advanced Computing and Mathematics Program Integrated Sensing and Processing (ISP) Initiative under  ARO
grant \#~DAAD19-02-1-0180, by NSF under grants \#~ECS-0225449
and~\#~CNS-0428404, by an IBM Faculty Award, and by  the Office of
Naval Research under MURI N000140710747.}}

\begin{document}
\maketitle \thispagestyle{empty} \maketitle

\begin{abstract}
The paper studies average consensus with random topologies (intermittent links)
 \emph{and} noisy channels. Consensus with noise in the network links
 leads to the bias-variance dilemma--running consensus for long
 reduces the bias of the final average estimate but increases its variance.
 We present two different compromises to this tradeoff:
 the $\mathcal{A-ND}$ algorithm modifies
 conventional consensus by forcing the weights to satisfy a
 \emph{persistence} condition (slowly decaying to zero;) and
the $\mathcal{A-NC}$ algorithm where the weights are constant but consensus is
 run for a fixed number of iterations~$\widehat{\imath}$,
then it is restarted and rerun for a total of~$\widehat{p}$ runs,
and at the end averages the final states of the $\widehat{p}$ runs
(Monte Carlo averaging). We use controlled Markov processes and
stochastic approximation arguments to prove almost sure
convergence of $\mathcal{A-ND}$ to the desired average (asymptotic
unbiasedness) and compute explicitly the m.s.e.~(variance) of the
consensus limit. We show that $\mathcal{A-ND}$ represents the best
of both worlds--low bias and low variance--at the cost of a slow
convergence rate; rescaling the weights balances the  variance
versus the rate of bias reduction (convergence rate).
 In contrast, $\mathcal{A-NC}$, because
of its constant weights, converges fast but presents a different
bias-variance tradeoff. For the same number of iterations
$\widehat{\imath}\widehat{p}$, shorter runs
(smaller~$\widehat{\imath}$) lead to high bias but smaller
variance (larger number $\widehat{p}$ of runs to average over.)
For a static non-random network with Gaussian noise, we compute
the optimal gain for $\mathcal{A-NC}$ to reach in the shortest run
length $\widehat{\imath}$,
 with high probability ($1-\delta$), $(\epsilon, \delta)$-consensus
 ($\epsilon$ residual bias).
Our results hold under fairly general assumptions on the random
link failures and communication noise.

\end{abstract}

\hspace{.43cm}\textbf{Keywords:} Consensus, random topology,
additive noise, sensor networks, stochastic approximation,
convergence

\newpage

\section{Introduction}
\label{introduction}Distributed computation in sensor networks  is a well-studied field with an extensive body of literature (see, for
example, \cite{tsitsiklisphd84} for early work.) Average consensus
computes iteratively the global average of distributed data using
local communications, see \cite{Olfati-2003,jadbabailinmorse03,reynolds87,vicsek95}
 that consider  versions and extensions of basic consensus.
  A review of the consensus literature is in~\cite{olfatisaberfaxmurray07}. 
  Reference~\cite{Boyd} designs the optimal link weights that optimize the convergence rate of the consensus algorithm when the connectivity graph of the network is fixed (not random). Our previous work, \cite{Allerton06-K-M,Asilomar06-K-M,tsp06-K-A-M,tsp07-K-M}, extends \cite{Boyd} by  designing the topology, i.e., both the weights and the connectivity graph, under a variety of conditions,
  including random links and link communication costs,
  under a network communication budget constraint.

  We consider distributed average consensus when
  \emph{simultaneously} the network topology is
  random (link failures, like when packets are
  lost in data networks) \emph{and} the communications among sensors is commonly
  noisy. A typical example is time division multiplexing, where,
  in a particular user's time slot the channel may not be
  available, and, if available, we assume the communication is
  analog and noisy.
  Our approach can handle spatially correlated link failures through Markovian
  sequences of Laplacians and certain types of Markovian noise, which go beyond
  independently, identically distributed (i.i.d.) Laplacian
  matrices and i.i.d. communication noise sequences.
  Noisy consensus leads to a tradeoff between bias and variance.
  Running consensus longer reduces bias, i.e., the error between
  the desired average and the consensus reached. But, due to noise,
  the variance of the limiting consensus
  grows with longer runs.  To address this dilemma, we consider two
  versions of consensus with link failures \emph{and}
  noise that represent two
  different bias-variance tradeoffs: the $\mathcal{A-ND}$
  and the $\mathcal{A-NC}$ algorithms.

$\mathcal{A-ND}$ updates each sensor state with a weighted fusion
of its current neighbors' states (received distorted by noise).
The fusion weights $\alpha(i)$ satisfy a \emph{persistence}
condition, decreasing to zero, but not too fast. $\mathcal{A-ND}$
falls under the purview of controlled Markov processes and we use
stochastic approximation techniques to  prove its almost
sure~(a.s.) consensus when the network is connected on the
average: the sensor state vector sequence converges a.s.~to the
\emph{consensus subspace}. A simple condition on the mean
Laplacian, $\overline{L}=\mathbb{E}\left\{L\right\}$, for
connectedness is on its second eigenvalue,
$\lambda_{2}\left(\overline{L}\right)>0$. We establish that the
sensor states converge asymptotically a.s.~to a finite random
variable~$\theta$ and, in particular, the expected sensor states
converge to the desired average~$r$ (asymptotic unbiasedness.) We
determine the variance of~$\theta$, which is the mean square error
(m.s.e.) between $\theta$ and the desired average. By properly
tuning the weights sequence $\{\alpha(i)\}$, the
variance~of~$\theta$ can be made arbitrarily small, though at a
cost of slowing $\mathcal{A-ND}$'s convergence rate, i.e., the
rate at which the bias goes to zero.


$\mathcal{A-NC}$ is a
repeated averaging algorithm that performs in-network
Monte-Carlo simulations: it runs consensus  $\widehat{p}$~times with
constant weight~$\alpha$, for a fixed number of iterations
$\widehat{\imath}$ each time and then each sensor averages its $\widehat{p}$
values of the state at the final iteration $\widehat{\imath}$ of each run.
 $\mathcal{A-NC}$'s constant weight $\alpha$ speeds its convergence rate relative to
$\mathcal{A-ND}$'s, whose weights $\alpha(i)$ decrease to
zero. We determine the number of iterations $\widehat{\imath}\widehat{p}$
required to reach $(\epsilon,\delta)$-\emph{consensus}, i.e., for the bias
of the consensus limit at each sensor
to be smaller than~$K\epsilon$, with high probability $(1-\delta)$. 
 For non-random networks, we establish a tight upper bound on the minimizing
 $\widehat{\imath}\widehat{p}$ and compute the corresponding optimal
 constant weight $\alpha$. We quantify the tradeoff between
the number of iterations $\widehat{\imath}$ per Monte-Carlo run
and the number of runs~$\widehat{p}$.

Finally, we compare the bias-variance tradeoffs between the two algorithms and
the network parameters that determine their convergence rate
and noise resilience. The fixed weight $\mathcal{A-NC}$ algorithm
can converge faster but requires greater inter-sensor coordination than
the $\mathcal{A-ND}$ algorithm.


\textbf{Comparison with existing literature.} Random link failures
and additive channel noise have been considered \emph{separately}.
Random link failures, but noise\emph{less} consensus, is
in~\cite{tsp07-K-M,Mesbahi,Bucklew,ChaiWu,Jadbabai, Porfiri}.
References~\cite{tsp07-K-M,Mesbahi,Bucklew} assume an erasure
model: the network links fail independently in space
(independently of each other) and in time (link failure events are
temporally independent.) Papers~\cite{ChaiWu,Porfiri} study
directed topologies with only time i.i.d.~link failures, but
impose distributional assumptions on the link formation process.
In~\cite{Jadbabai} the link failures are i.i.d.~Laplacian
matrices, the graph is directed, and no distributional assumptions
are made on the Laplacian matrices. The paper presents necessary
and sufficient conditions for consensus using the ergodicity of
products of stochastic matrices.

Similarly, \cite{Mesbahi-Noise,Huangmanton-acc07,Huang}
consider consensus with additive noise,
but \emph{fixed} or static, \emph{non random} topologies
(no link failures.) They use
a decreasing weight sequence to guarantee consensus.
These references do not characterize the m.s.e. For example,
\cite{Huangmanton-acc07,Huang} rely on the existence of a unique
solution to an algebraic Lyapunov equation.
The more  general problem of distributed estimation (of which
average consensus is a special case) in the presence of additive
noise is in~\cite{Giannakis-Estimation}, again with a fixed
topology. Both~\cite{Mesbahi-Noise,Giannakis-Estimation} assume a
temporally white noise sequence, while our approach can
accommodate a more general Markovian noise sequence, in addition
to white noise processes.

In summary, with respect to
 \cite{tsp07-K-M}--\cite{Giannakis-Estimation}, our approach considers:
  \begin{inparaenum}[i)] \item random
topologies \emph{and} noisy communication links \emph{simultaneously};
\item spatially correlated (Markovian) dependent random link failures;
\item time Markovian noise sequences; \item undirected topologies;
\item no distributional assumptions; \item consensus (estimation
being considered elsewhere;) and \item two versions of consensus representing
different compromises of bias versus variance.
\end{inparaenum}

Briefly, the paper is as follows.  Sections~\ref{graphtheory} and~\ref{distavgcons}
summarize relevant spectral graph and
average consensus results. Sections~\ref{A-NDProbFormAssumptions}
and~\ref{ConsRepAvg} treat the additive noise \emph{with} random link failure communication analyzing the $\mathcal{A-ND}$ and
$\mathcal{A-NC}$ algorithms, respectively.
Finally, Section~\ref{conclusion} concludes the paper. 

\section{Elementary Spectral Graph Theory}
\label{graphtheory}
We summarize briefly facts from spectral graph
theory. For an undirected graph $G=(V,E)$, $V=\left[1\cdots
N\right]$ is the set of nodes or vertices, $|V|=N$, and $E$ is the
set of edges, $|E|=M$. The unordered pair $(n,l)\in E$ if there
exists an edge between nodes $n$ and $l$. We only consider simple
graphs, i.e., graphs devoid of self-loops and multiple edges. A path between
nodes $n$ and $l$ of length $m$ is a sequence
$(n=i_{0},i_{1},\cdots,i_{m}=l)$ of vertices, such that,
$(i_{k},i_{k+1})\in E,\,  0\leq k\leq m-1$. A
graph is connected if there exists a path, between each
pair of nodes. The neighborhood of node~$n$ is
\begin{equation}
\label{def:omega} \Omega_{n}=\left\{l\in V~|~(n,l)\in
E\right\} 
\end{equation}
Node $n$ has degree $d_{n}=|\Omega_{n}|$ (number of edges with $n$ as one end point.) The structure of the graph can be described by the symmetric $N\times N$ adjacency matrix, $A=\left[A_{nl}\right]$, $A_{nl}=1$, if $(n,l)\in E$, $0$~otherwise.
 Let the degree matrix  be the diagonal matrix $D=\mbox{diag}\left(d_{1}\cdots
d_{N}\right)$. The graph Laplacian matrix, $L$, is
\begin{equation}
\label{def_L} L=D-A
\end{equation}
 The Laplacian is a positive semidefinite matrix; hence,
its eigenvalues can be ordered as
\begin{equation}
\label{def_L_eig}
0=\lambda_{1}(L)\leq\lambda_{2}(L)\leq\cdots\leq\lambda_{N}(L)
\end{equation}
The multiplicity of the zero eigenvalue equals the number of
connected components of the network; for a
connected graph, $\lambda_{2}(L)>0$. This second eigenvalue is the algebraic connectivity or the Fiedler value of the network; see \cite{FanChung,Mohar,SensNets:Bollobas98} for detailed
treatment of graphs and their spectral theory.

\section{Distributed Average Consensus with Imperfect Communication}
\label{distavgcons} In a simple form, distributed average
consensus computes the average $r$ of the initial node data
\begin{equation}
\label{def_r} r=\frac{1}{N}\sum_{n=1}^{N}x_{n}(0)
\end{equation}
by local data exchanges among neighbors. For noiseless and
unquantized data exchanges across the network links, the state of
each node is updated iteratively by
\begin{equation}
\label{cons_state_update} x_{n}(i+1)=w_{nn}(i)x_{n}(i)+
\sum_{l\in\Omega_{n}(i)}w_{nl}(i)x_{l}(i),~~~1\leq n\leq N
\end{equation}
where the link weights, $w_{nl}$'s, may be constant or time
varying. Similarly, the topology of a time-varying network is
captured by making the neighborhoods, $\Omega_{n}$'s, to be a
function of time. Because noise causes consensus to diverge, \cite{SensNets:Xiao05,tsp06-K-A-M},
we let the link weights to be the same across different
network links, but vary with time.
 Eq.~(\ref{cons_state_update}) becomes 
\begin{equation}
\label{eqweightsupdate}
x_{n}(i+1)=\left[1-\alpha(i)d_{n}(i)\right]x_{n}(i)+
\alpha(i)\sum_{l\in\Omega_{n}(i)}x_{l}(i), \:\: 1\leq n\leq N
\end{equation}
We address consensus with imperfect inter-sensor communication,
where each sensor receives noise corrupted versions of its
neighbors' states. Eq.~(\ref{eqweightsupdate}) is now This leads
to the state update given by
\begin{equation}
\label{genstateupdate}
x_{n}(i+1)=\left[1-\alpha(i)d_{n}(i)\right]x_{n}(i)+
\alpha(i)\sum_{l\in\Omega_{n}(i)}f_{nl,i}\left[x_{l}(i)\right],
\:\: 1\leq n\leq N
\end{equation}
where $\{f_{nl,i}\}_{1\leq n,l\leq N,~i\geq 0}$ is a sequence of
functions (possibly random) modeling the channel imperfections. In
the following sections, we analyze the consensus problem given by
eqn.~(\ref{genstateupdate}), when the channel communication is
corrupted by additive noise. In~\cite{karmoura-quantized}, we
consider the effects of quantization (see also~\cite{Tuncer} for a
treatment of consensus algorithms with quantized communication.)
Here, we study two different algorithms. The first,
$\mathcal{A-ND}$ , considers a decreasing weight sequence
($\alpha(i)\rightarrow 0$) and is analyzed in
Section~\ref{A-NDProbFormAssumptions}. The second,
$\mathcal{A-NC}$, uses  repeated averaging with a constant link
weight and is detailed in Section~\ref{ConsRepAvg}.

\section{$\mathcal{A-ND}$: Consensus in Additive Noise and Random Link Failures}
\label{A-NDProbFormAssumptions}We consider
distributed consensus when the network links fail or become alive at
random times, and data exchanges are corrupted by additive noise.
The network topology varies randomly
across iterations. We analyze the convergence properties of the $\mathcal{A-ND}$
algorithm under this generic scenario. We start by formalizing the assumptions underlying $\mathcal{A-ND}$ in the next Subsection.

\subsection{\textbf{Problem Formulation and Assumptions}}
\label{ProbFormStochApp}
 We compute the average of the initial state $\mathbf{x}(0)=\left[x_{1}(0)\cdots x_{N}(0)\right]^{T}\in\mathbb{R}^{N\times 1}$ with the distributed consensus algorithm with communication channel imperfections given in
eqn.~(\ref{genstateupdate}). Let $\{v_{nl}(i)\}_{1\leq n,l\leq N,~i\geq 0}$ be a sequence of independent zero mean random variables. For additive noise,
\begin{equation}
\label{def_f_nli} f_{nl,i}(y)=y+v_{nl}(i)
\end{equation}
Recall the Laplacian~$L$ defined in~(\ref{def_L}). Collecting the states
$x_n(i)$ in the vector $\mathbf{x}(i)$, eqn.~(\ref{genstateupdate}) is
\begin{eqnarray}
\label{A-NDmain}
\mathbf{x}(i+1)&=&\mathbf{x}(i)-\alpha(i)\left[L(i)\mathbf{x}(i)+\mathbf{n}(i)\right]\\
\label{A-NDmain1}
\left[\mathbf{n}(i)\right]_l&=&n_{l}(i)=-\sum_{k\in\Omega_{l}(i)}v_{lk}(i),\:1\leq l\leq N, \:i\geq
0
\end{eqnarray}
We now state the assumptions of the $\mathcal{A-ND}$
algorithm.\footnote{See also~\cite{Asilomar07-K-M}, where parts of
the results are presented.}

\begin{itemize}[\setlabelwidth{1)}]

\item{\textbf{1)}} \textbf{Random Network Failure}: We propose two
models; the second is more general than the first.

\begin{itemize}[\setlabelwidth{1.1)}]

\item{\textbf{1.1)}} \textbf{Temporally i.i.d.~Laplacian Matrices}: The graph
Laplacians are
\begin{equation}
\label{Lcond} L(i) = \overline{L}+\widetilde{L}(i),~\forall i\geq
0
\end{equation}
where $\{L(i)\}_{i\geq 0}$ is a sequence of i.i.d.~Laplacian
matrices with mean $\overline{L} = \mathbb{E}\left[L(i)\right]$,
such that $\lambda_{2}\left(\overline{L}\right)>0$. We do not make
any distributional assumptions on the link failure model, and, in
fact, as long as the sequence $\{L(i)\}_{i\geq 0}$ is independent
with constant mean $\overline{L}$, satisfying
$\lambda_{2}\left(\overline{L}\right)>0$, the i.i.d. assumption
can be dropped. During the same iteration, the link failures can
be spatially dependent, i.e., correlated across different edges of
the network. This model subsumes the erasure network model, where
the link failures are independent both over space and time.
Wireless sensor networks motivate this model since interference
among the sensors communication correlates the link failures over
space, while over time, it is still reasonable to assume that the
channels are memoryless or independent.

Connectedness of the graph is an important issue. We do not
require that the random instantiations~$G(i)$ of the graph be
connected; in fact, it is possible to have all these
instantiations to be disconnected. We only require that the graph
stays connected on \emph{average}. This is captured by requiring
that $\lambda_{2} \left(\overline{L}\right)>0$, enabling us to
capture a broad class of asynchronous communication models; for
example, the random asynchronous gossip protocol analyzed
in~\cite{Boyd-Gossip} satisfies $\lambda_{2}\left(\overline{L}
\right)>0$ and hence falls under this framework.

\item{\textbf{1.2)}} \textbf{Temporally Markovian Laplacian Matrices}: Our results hold when the Laplacian matrix sequence $\{L(i,\mathbf{x}(i))\}_{i\geq 0}$ is state-dependent.
%
 More precisely, we assume that there exists a two-parameter random
field, $\{L(i,\mathbf{x})\}_{i\geq
0,\mathbf{x}\in\mathbb{R}^{N\times 1}}$ of Laplacian matrices such
that
\begin{equation}
\label{Lcond1} \mathbb{E}[L(i,\mathbf{x})]=\overline{L},~~\forall
i,\mathbf{x}
\end{equation}
and $\lambda_{2}(\overline{L})>0$. We also require that, for a
fixed $i$, the random matrices,
$\{L(i,\mathbf{x})\}_{\mathbf{x}\in\mathbb{R}^{N\times 1}}$, are
independent of the sigma algebra, $\sigma\left(\mathbf{x}(j),
0\leq j\leq i\right)$.\footnote{This guarantees that the Laplacian
$L(i,\mathbf{x}(i))$ may depend on the past state history
$\{\mathbf{x}(j),~j\leq i\}$, only through the present state
$\mathbf{x}(i)$.} It is clear then that the Laplacian matrix
sequence, $\{L(i,\mathbf{x}(i))\}_{i\geq 0}$, is Markov. We will
show that our convergence analysis holds also for this general
link failure model.  Such a model may be appropriate in stochastic
formation control scenarios,
see~\cite{Olfati_flocking,Jadbabaie_flocking,Mesbahi-StateLaplacian},
where the network topology is state-dependent.
%
\end{itemize}
%
%
%
\item{\textbf{2)}} \textbf{Communication Noise Model}: We propose two models; the second is more general than the first.
\begin{itemize}[\setlabelwidth{1.1)}]

\item{\textbf{2.1)}} \textbf{Independent Noise Sequence}: The
additive noise $\{v_{nl}(i)\}_{1\leq n,l\leq N,~i\geq 0}$ is an
independent sequence
\begin{equation}
\label{vnlcond} \mathbb{E}\left[v_{nl}(i)\right] = 0,\:\:\forall
1\leq n,l\leq N,\:\:i\geq
0,\:\:\sup_{n,l,i}\mathbb{E}\left[v_{nl}^{2}(i)\right]=\mu <\infty
\end{equation}
The sequences, $\{L(i)\}_{i\geq 0}$ and
$\{v_{nl}(i)\}_{1\leq n,l\leq N,~i\geq 0}$ are mutually
independent. Hence, $L(i)$,
$\{v_{nl}(i)\}$, ${1\leq n,l\leq N}$, $i\geq 0$ are independent of $\sigma\left(\mathbf{x}(j), 0\leq j\leq i\right)$, $\forall i$. Then, from eqn.~(\ref{A-NDmain1}),
\begin{equation}
\label{ncond}\mathbb{E}\left[\mathbf{n}(i)\right]=
\mathbf{0},\:\:\forall i,\:\:
\sup_{i}\mathbb{E}\left[\|\mathbf{n}(i)\|^{2}\right]=\eta\leq
N(N-1)\mu <\infty
\end{equation}
No distributional assumptions are required on the noise sequence.

\item{\textbf{2.2)}} \textbf{Markovian Noise Sequence}: Our
approach allows the noise sequence to be Markovian through
state-dependence. Let the two-parameter random field,
$\{\mathbf{n}(i,\mathbf{x})\}_{i\geq
0,\mathbf{x}\in\mathbb{R}^{N\times 1}}$ of random vectors
\begin{equation}
\label{ncond2} \mathbb{E}[\mathbf{n}(i,\mathbf{x})]=0,\:\forall
i,\mathbf{x}
\end{equation}
For fixed $i$, the random vectors,
$\{\mathbf{n}(i,\mathbf{x})\}_{\mathbf{x}\in\mathbb{R}^{N\times
1}}$, are independent of the $\sigma$-algebra,
$\sigma\left(\mathbf{x}(j), 0\leq j\leq i\right)$ and
 the random families
$\{L(i,\mathbf{x})\}_{\mathbf{x}\in\mathbb{R}^{N\times 1}}$ and
$\{\mathbf{n}(i,\mathbf{x})\}_{\mathbf{x}\in\mathbb{R}^{N\times
1}}$ are independent. It is clear then that the noise vector sequence,
$\{\mathbf{n}(i,\mathbf{x}(i))\}_{i\geq 0}$, is Markov.
Note, however, in this case the resulting Laplacian and noise sequences,
$\left\{L(i,\mathbf{x}(i))\right\}_{i\geq0}$ and
$\left\{\mathbf{n}(i,\mathbf{x}(i))\right\}_{i\geq0}$ are no longer
independent; they are coupled through the state $\mathbf{x}(i)$.
In addition to~(\ref{ncond2}), we require
the variance of the noise component orthogonal to the consensus
subspace (see eqn.~(\ref{ConsSubspace})) to satisfy,
for constants, $c_{1},c_{2}\geq 0$,
\begin{equation}
\label{ncond3}
\mathbb{E}[\|\mathbf{n}_{\mathcal{C}^{\perp}}(i,\mathbf{x})\|^{2}]\leq
c_{1}+c_{2}\|\mathbf{x}_{\mathcal{C}^{\perp}}\|^{2}
\end{equation}
We do not restrict the variance growth rate of the noise component in
the consensus subspace. This clearly subsumes the bounded noise
variance model. An example of such noise is
\begin{equation}
\label{ncond4}
\mathbf{n}(i,\mathbf{x}(i))=\vartheta(i)\left(\mathbf{x}(i)+\mathbf{w}(i)\right)
\end{equation}
where $\{\vartheta(i)\}_{i\geq 0}$ and $\{\mathbf{w}(i)\}_{i\geq
0}$ are zero mean finite variance mutually i.i.d.~sequences of
scalars and vectors, respectively. It is then clear that the
condition in eqn.~(\ref{ncond3}) is satisfied, and the noise
model~\textbf{2.2)} applies. The model in eqn.~(\ref{ncond4})
arises, for example, in multipath effects in MIMO systems, when
the channel adds multiplicative noise whose amplitude is
proportional to the transmitted data.
\end{itemize}
%
%
\item{\textbf{3)}} \textbf{Persistence Condition}: The weights
decay to zero, but not too fast
\begin{equation}
\label{alphacond} \alpha (i)>0,\:\:\sum_{i\geq 0}\alpha
(i)=\infty,\:\:\sum_{i\geq 0}\alpha^{2}(i)<\infty
\end{equation}
This condition is commonly assumed in adaptive control and
signal processing. Examples include
\begin{equation}
\label{eq:alphai} \alpha(i)=\frac{1}{i^\beta},\:\: .5<\beta\leq 1
\end{equation}
\end{itemize}
For clarity, in the main body of the paper, we prove the results for the
$\mathcal{A-ND}$ algorithm under
Assumptions~\textbf{1.1)}, \textbf{2.1)}, and \textbf{3)}. In the Appendix, we point out how to modify the proofs when the more general assumptions~\textbf{1.2)} and \textbf{2.2)} hold.

We now prove the almost sure~(a.s.) convergence of the
$\mathcal{A-ND}$ algorithm in eqn.(\ref{A-NDmain}) by using
results from the theory of stochastic approximation algorithms,
\cite{Nevelson}.

\subsection{\textbf{A Result on Convergence of Markov Processes}}
\label{ConvMarkovProc} A systematic and thorough treatment of
stochastic approximation procedures has been given
in~\cite{Nevelson}. In this section, we modify slightly  a result
from~\cite{Nevelson} and restate it as a theorem in a form
relevant to our application. We follow the notation
of~\cite{Nevelson}, which we now introduce.

Let $\mathbf{X}=\left\{\mathbf{x}(i)\right\}_{i\geq 0}$ be a
Markov process on $\mathbb{R}^{N\times 1}$. The generating
operator $\mathcal{L}$ of $\mathbf{X}$ is
\begin{equation}
\label{Lgenop}
\mathcal{L}V(i,\mathbf{x})=
\mathbb{E}\left[V\left(i+1,\mathbf{x}(i+1)\right)|\mathbf{x}(i)=\mathbf{x}\right]-V(i,\mathbf{x})
\end{equation}
for functions $V(i,\mathbf{x}),~i\geq
0,~\mathbf{x}\in\mathbb{R}^{N\times 1}$, provided the conditional
expectation exists. We say that $V(i,\mathbf{x})\in
D_{\mathcal{L}}$ in a domain $A$, if $\mathcal{L}V(i,\mathbf{x})$
is finite for all $(i,\mathbf{x})\in A$.

Denote the Euclidean metric by $\rho(\cdot)$. For
$B\subset\mathbb{R}^{N\times 1}$,  the $\epsilon$-neighborhood of
$B$ and its complement is
\begin{eqnarray}
\label{Uep} U_{\epsilon}(B) &=& \left\{x~|~\inf_{y\in B}\rho
(x,y)<\epsilon\right\}\\
\label{def_Vep} V_{\epsilon}(B) &=& \mathbb{R}^{N\times 1}\backslash
U_{\epsilon}(B)
\end{eqnarray}

We now state the desired theorem, whose proof we sketch in the Appendix.
\begin{theorem}
\label{MarkovProc} Let $\mathbf{X}$ be a Markov process with
generating operator $\mathcal{L}$. Let there exist a non-negative
function $V(i,\mathbf{x})\in D_{\mathcal{L}}$ in the domain $i\geq
0,~\mathbf{x}\in\mathbb{R}^{N\times 1}$, and
$B\subset\mathbb{R}^{N\times 1}$ with the following properties:
%
\begin{eqnarray}
\label{Vcond1} \hspace{-1in}\mbox{\textbf{1)}}
\hspace{2in}\inf_{i\geq 0,\mathbf{x}\in
V_{\epsilon}(B)}V(i,\mathbf{x})&>&0,\:\:\forall \epsilon> 0\hspace{1in}\\
\label{Vcond2} V(i,\mathbf{x})&\equiv& 0,\:\:\mathbf{x}\in B\\
\label{Vcond3} \lim_{\mathbf{x}\rightarrow B}\sup_{i\geq 0}
V(i,\mathbf{x})&=&0\\
%
%
\label{LVcond1} \hspace{-.6in}\mbox{\textbf{2)}}\hspace{+.6in}
\hspace{2in}\mathcal{L}V(i,\mathbf{x})&\leq&
g(i)(1+V(i,\mathbf{x}))-\alpha (i)\varphi (i,\mathbf{x})
\end{eqnarray}
where $\varphi(i,\mathbf{x}),i\geq
0,~\mathbf{x}\in\mathbb{R}^{N\times 1}$ is a non-negative function
such that
\begin{equation}
\label{LVcond2} \inf_{i,\mathbf{x}\in V_{\epsilon}(B)}\varphi
(i,\mathbf{x})
>0,\:\:\forall \epsilon >0
\end{equation}
%
%
\begin{eqnarray}
\label{alphacond-1}
\hspace{-2.25in}\mbox{\textbf{3)}} \hspace{2.25in}\alpha(i) &>&0,\:\:\sum_{i\geq 0}\alpha (i)= \infty\\
\label{gcond} g(i)&>&0,\:\:\sum_{i\geq 0}g(i)<\infty
\end{eqnarray}
Then, the Markov process
$\mathbf{X}=\left\{\mathbf{x}(i)\right\}_{i\geq 0}$ with arbitrary
initial distribution converges a.s. to $B$ as
$i\rightarrow\infty$. In other words,
\begin{equation}
\label{convmarkov12}
\mathbb{P}\left(\lim_{i\rightarrow\infty}\rho\left(\mathbf{x}(i),B\right)=0\right)=1
\end{equation}
\end{theorem}

\subsection{\textbf{Proof of Convergence of the $\mathcal{A-ND}$ Algorithm}}
\label{ConvproofAND} The $\mathcal{A-ND}$ distributed
consensus algorithm
  is given by eqn.~(\ref{A-NDmain}) in Section~\ref{ProbFormStochApp}. To establish its a.s.~convergence using Theorem~\ref{MarkovProc}, define the \emph{consensus} subspace, $\mathcal{C}$, aligned with~$\mathbf{1}$, the vectors of~$1$'s,
\begin{equation}
\label{ConsSubspace} \mathcal{C} = \left\{\mathbf{x}\in\mathbb{R}^{N\times
1}\,\left|\right.\,\mathbf{x}=a\mathbf{1}, \,a\in\mathbb{R}\right\}
\end{equation}
We recall a result on distance properties in $\mathbb{R}^{N\times
1}$ to be used in the sequel. We omit the proof.
\begin{lemma}
\label{distprop} Let $\mathcal{S}$ be a subspace of
$\mathbb{R}^{N\times1}$. For $\mathbf{x}\in\mathbb{R}^{N\times 1}$, consider the orthogonal decomposition
$\mathbf{x}= \mathbf{x}_{\mathcal{S}}+\mathbf{x}_{\mathcal{S}^{\perp}}$.
Then
$\rho(\mathbf{x},\mathcal{S})=\left\|\mathbf{x}_{\mathcal{S}^{\perp}}\right\|$.
\end{lemma}
%
\begin{theorem}[$\mathcal{A-ND}$ a.s. convergence]
\label{ConvProof} Let assumptions~\textbf{1.1)},
\textbf{2.1)}, and \textbf{3)} hold. Consider the $\mathcal{A-ND}$
consensus algorithm in eqn.~(\ref{A-NDmain}) in
Section~\ref{ProbFormStochApp} with initial state
$\mathbf{x}(0)\in\mathbb{R}^{N\times 1}$. Then,
\begin{equation}
\label{th-2.1}
\mathbb{P}\left[\lim_{i\rightarrow\infty}\rho(\mathbf{x}(i),\mathcal{C})=0\right]=1
\end{equation}
\end{theorem}
\begin{proof}
Under the assumptions, the process
$\mathbf{X}=\left\{\mathbf{x}(i)\right\}_{i\geq 0}$ is Markov.
Define
\begin{equation}
\label{defV} V(i,\mathbf{x})= \mathbf{x}^{T}\overline{L}\mathbf{x}
\end{equation}
The potential function $V(i,\mathbf{x})$ is non-negative. Since $\mathbf{x}\in \mathcal{C}$ is an eigenvector of~$\overline{L}$ with zero eigenvalue,
\begin{equation}
\label{assV1} V(i,\mathbf{x})\equiv
0,\:\mathbf{x}\in\mathcal{C}, \:\:\:\lim_{\mathbf{x}\rightarrow
\mathcal{C}}\sup_{i\geq 0} V(i,\mathbf{x})=0
\end{equation}
The second condition follows from the continuity of
$V(i,\mathbf{x})$. By Lemma~\ref{distprop} and the definition
in eqn.~(\ref{def_Vep}) of the complement of the $\epsilon$-neighborhood of a set
\begin{equation}
\label{assV2.1} \mathbf{x}\in
V_{\epsilon}(\mathcal{C})~\Longrightarrow ~
\|\mathbf{x}_{\mathcal{C}^{\perp}}\|\geq\epsilon
\end{equation}
Hence, for $\mathbf{x}\in V_{\epsilon}(\mathcal{C})$,
\vspace*{-.5cm}
\begin{eqnarray}
\label{assV2.2} V(i,\mathbf{x}) & = & \mathbf{x}^{T}\overline{L}\mathbf{x} \\
\nonumber
                       & \geq &
                       \lambda_{2}\left(\overline{L}\right)\|\mathbf{x}_{\mathcal{C}^{\perp}}\|^{2}\\
                       \nonumber
                       & \geq & \lambda_{2}\left(\overline{L}\right)\epsilon^{2}
\end{eqnarray}
Then, since by assumption \textbf{1.1)}  $\lambda_{2}\left(\overline{L}\right)>0$ (note that
the assumption $\lambda_{2}\left(\overline{L}\right)>0$ comes into
play here), we get
\begin{equation}
\label{assV2.3} \inf_{i\geq 0,\mathbf{x}\in
V_{\epsilon}(\mathcal{C})}V(i,\mathbf{x})\geq
\lambda_{2}\left(\overline{L}\right)\epsilon^{2}
>0
\end{equation}
Now consider $\mathcal{L}V(i,\mathbf{x})$ in eqn.~(\ref{Lgenop}).
Using eqn.~(\ref{A-NDmain}) in~(\ref{Lgenop}), we obtain
\begin{eqnarray}
\label{assV3} \mathcal{L}V(i,\mathbf{x}) & = &
\mathbb{E}\left[\mathbf{x}(i+1)^{T}\overline{L}\mathbf{x}(i+1)~|~\mathbf{x}(i)=\mathbf{x}\right]-\mathbf{x}^{T}\overline{L}\mathbf{x}
 \\ & = &
\mathbb{E}\left[\left[\mathbf{x}-\alpha(i)\overline{L}\mathbf{x}-\alpha(i)\widetilde{L}(i)\mathbf{x}-\alpha(i)\mathbf{n}(i)\right]^{T}
\overline{L}\left[\mathbf{x}-\alpha(i)\overline{L}\mathbf{x}-\alpha(i)\widetilde{L}(i)\mathbf{x}-\alpha(i)\mathbf{n}(i)\right]\right]
\nonumber\\
&&-\mathbf{x}^{T}\overline{L}\mathbf{x}\nonumber
\end{eqnarray}
 Using the independence of $\widetilde{L}(i)$ and
$\mathbf{n}(i)$ with respect to $\mathbf{x}(i)$, that
$\widetilde{L}(i)$ and $\mathbf{n}(i)$ are zero-mean, and that the
subspace $\mathcal{C}$ lies in the null space of
$\overline{L}^{3}$ and $\widetilde{L}(i)$ (the latter because this is true for both $L(i)$ and $\overline{L}$), eqn.~(\ref{assV3})
leads successively to
\begin{eqnarray}
\label{assV3.0} \mathcal{L}V(i,\mathbf{x}) & = &
-2\alpha(i)\mathbf{x}^{T}\overline{L}^{2}\mathbf{x}+
\alpha^{2}(i)\mathbf{x}^{T}\overline{L}^{3}\mathbf{x}
+\mathbb{E}\left[\alpha^{2}(i)\left(\widetilde{L}(i)\mathbf{x}\right)^{T}\overline{L}
\left(\widetilde{L}(i)\mathbf{x}\right)\right]
\hspace{-.075cm}+\hspace{-.075cm}\mathbb{E}\left[\alpha^{2}(i)\mathbf{n}(i)^{T}\overline{L}\mathbf{n}(i)\right]
\nonumber \\
& \leq & -2\alpha(i)\mathbf{x}^{T}\overline{L}^{2}\mathbf{x}
+\alpha^{2}(i)\lambda_{N}(\overline{L})^{3}\|\mathbf{x}_{\mathcal{C}^{\perp}}\|^{2}
+\alpha^{2}(i)\lambda_{N}(\overline{L})\mathbb{E}\left[\|\widetilde{L}(i)\mathbf{x}\|^{2}\right]
\nonumber\\
&&+\alpha^{2}(i)\lambda_{N}(\overline{L})\mathbb{E}\left[\|\mathbf{n}(i)\|^{2}\right]
\nonumber \\ & \leq &
-2\alpha(i)\mathbf{x}^{T}\overline{L}^{2}\mathbf{x}
+\alpha^{2}(i)\lambda_{N}(\overline{L})^{3}\|\mathbf{x}_{\mathcal{C}^{\perp}}\|^{2}
+\alpha^{2}(i)\lambda_{N}(\overline{L})\mathbb{E}\left[\lambda_{\mbox{\scriptsize{max}}}^{2}
\left(\widetilde{L}(i)\right)\right]\|\mathbf{x}_{\mathcal{C}^{\perp}}\|^{2}
\nonumber\\
&&+\alpha^{2}(i)\lambda_{N}(\overline{L})\eta
\nonumber \\
 & \leq &
-2\alpha(i)\mathbf{x}^{T}\overline{L}^{2}\mathbf{x}+\alpha^{2}(i)\lambda_{N}(\overline{L})^{3}\|\mathbf{x}_{\mathcal{C}^{\perp}}\|^{2}+4\alpha^{2}(i)N^{2}\lambda_{N}(\overline{L})\|\mathbf{x}_{\mathcal{C}^{\perp}}\|^{2}+\alpha^{2}(i)\lambda_{N}(\overline{L})\eta
\end{eqnarray}
The last step follows because all the eigenvalues
of $\widetilde{L}(i)$ are less than $2N$ in absolute value, by the
Gershgorin circle theorem. Now, by the fact
$\mathbf{x}^{T}\overline{L}\mathbf{x}\geq\lambda_{2}\left(\overline{L}\right)\|\mathbf{x}_{\mathcal{C}^{\perp}}\|^{2}$
and $\lambda_{2}(\overline{L})>0$, we have
\begin{eqnarray}
\label{assV3.1} \mathcal{L}V(i,\mathbf{x}) & \leq &
-2\alpha(i)\mathbf{x}^{T}\overline{L}^{2}\mathbf{x} +
\alpha^{2}(i)\left[\lambda_{N}(\overline{L})\eta+
\frac{\lambda_{N}^{3}(\overline{L})}{\lambda_{2}\left(\overline{L}\right)}\mathbf{x}^{T}\overline{L}\mathbf{x}
+\frac{4N^{2}\lambda_{N}(\overline{L})}{\lambda_{2}\left(\overline{L}\right)}\mathbf{x}^{T}\overline{L}\mathbf{x}\right]
\nonumber \\ & \leq &
-\alpha(i)\varphi(i,\mathbf{x})+g(i)\left[1+V(i,\mathbf{x})\right]
\end{eqnarray}
where
\begin{equation}
\label{assV3.2}
\varphi(i,\mathbf{x})=2\mathbf{x}^{T}\overline{L}^{2}\mathbf{x},\:\:
g(i)=\alpha^{2}(i)\max\left(\lambda_{N}(\overline{L})\eta,\frac{\lambda_{N}^{3}(\overline{L})}{\lambda_{2}\left(\overline{L}\right)}
+\frac{4N^{2}\lambda_{N}(\overline{L})}{\lambda_{2}\left(\overline{L}\right)}\right)
\end{equation}
It is easy to see that $\varphi(i,\mathbf{x})$ and $g(i)$ defined
above satisfy the conditions for Theorem~\ref{MarkovProc}. Hence,
\begin{equation}
\label{finres}
\mathbb{P}\left[\lim_{i\rightarrow\infty}\rho(\mathbf{x}(i),\mathcal{C})=0\right]=1
\end{equation}
\end{proof}
Theorem~\ref{ConvProof} assumes~\textbf{1.1)}, \textbf{2.1)}, \textbf{3)}. For an equivalent statement under~\textbf{1.2)}, \textbf{2.2)}, see
Theorem~\ref{ConvProofgen} in the Appendix.

Theorem~\ref{ConvProof} shows that the sample paths approach the
consensus subspace, $\mathcal{C}$, with probability 1 as
$i\rightarrow\infty$. We now show that the sample paths, in fact,
converge a.s. to a finite point in $\mathcal{C}$.
\begin{theorem}[a.s.~consensus: Limiting random variable]
\label{ConvProof1} Let assumptions~\textbf{1.1)}, \textbf{2.1)}, and  \textbf{3)}
hold. Consider the $\mathcal{A-ND}$ consensus algorithm in
eqn.~(\ref{A-NDmain}) in Section~\ref{ProbFormStochApp} with
initial state $\mathbf{x}(0)\in\mathbb{R}^{N\times 1}$. Then,
there exists an almost sure finite real random variable
$\mathbf{\theta}$ such that
\begin{equation}
\label{th2.1}
\mathbb{P}\left[\lim_{i\rightarrow\infty}\mathbf{x}(i)=\mathbf{\theta}\mathbf{1}\right]
= 1
\end{equation}
\end{theorem}
\begin{proof}
Denote the average of $\mathbf{x}(i)$ by
\begin{equation}
\label{th2.2}
x_{\mbox{\scriptsize{avg}}}(i)=\frac{1}{N}\mathbf{1}^{T}\mathbf{x}(i)
\end{equation}
The conclusions of Theorem~\ref{ConvProof} imply that
\begin{equation}
\label{th2.3}
\mathbb{P}\left[\lim_{i\rightarrow\infty}\left\|\mathbf{x}(i)
-x_{\mbox{\scriptsize{avg}}}(i)\mathbf{1}\right\|=0\right]=1
\end{equation}
Recall the distributed average consensus algorithm in
eqn.~(\ref{A-NDmain}).
Premultiplying both sides of eqn.(\ref{A-NDmain}) by
$\frac{1}{N}\mathbf{1}^{T}$, we get the stochastic difference
equation for the averages as
\begin{eqnarray}
\label{th2.5} x_{\mbox{\scriptsize{avg}}}(i+1) & = &
x_{\mbox{\scriptsize{avg}}}(i)-\mathbf{\xi}(i)\\
\nonumber & = & x_{\mbox{\scriptsize{avg}}}(0)-\sum_{0\leq j\leq
i}\mathbf{\xi}(j)
\end{eqnarray}
where
\[
\mathbf{\xi}(i)=\frac{\alpha(i)}{N}\mathbf{1}^{T}\mathbf{n}(i)
\]
Given eqn.~(\ref{ncond}), in particular, the sequence
$\left\{\mathbf{n}(i)\right\}$ is time independent,
 it follows that
\begin{eqnarray}
\label{th2.6} \mathbb{E}\left[\mathbf{\xi}(i)\right] & = &
0,~\forall i
\nonumber \\
\sum_{i\geq 0}\mathbb{E}[\mathbf{\xi}(i)]^{2} & = & \sum_{i\geq
0}\frac{\alpha^{2}(i)}{N^{2}}\mathbb{E}\left[\|\mathbf{n}(i)\|^{2}\right]\\
\nonumber & \leq & \frac{\eta}{N^{2}}\sum_{i\geq 0}\alpha^{2}(i)\\
\nonumber & < & \infty
\end{eqnarray}
which implies
\begin{equation}
\label{th2.7}
\mathbb{E}\left[\left(x_{\mbox{\scriptsize{avg}}}(i)\right)^{2}\right]\leq
x_{\mbox{\scriptsize{avg}}}^{2}(0)+ \frac{\eta}{N^{2}}\sum_{j\geq
0}\alpha^{2}(j),~\forall i
\end{equation}
Thus, the sequence $\{x_{\mbox{\scriptsize{avg}}}(i)\}_{i\geq 0}$
is an $\mathcal{L}_{2}$ bounded martingale and, hence, converges
a.s. and in $\mathcal{L}_{2}$ to a finite random variable $\theta$
(see,~\cite{Williams}.) The theorem then follows from
eqn.~(\ref{th2.3}).
\end{proof}
Again, we note that we obtain an equivalent statement under Assumptions~\textbf{1.2)}, \textbf{2.1)} in Theorem~\ref{ConvProofGen2} in the Appendix. Proving under Assumption~\textbf{2.2)} requires more specific information about the mixing properties of the Markovian noise sequence, which
due to space limitations is not addressed here.


\subsection{\textbf{Mean Square Error}} \label{error}
 By Theorem~\ref{ConvProof1},  the sensors reach consensus
asymptotically and converge a.s.~to a finite random
variable $\theta$. Viewed as an estimate of the initial
average $r$ (see eqn.~(\ref{def_r})),  $\theta$ should
possess desirable properties like unbiasedness and small
m.s.e. The next Lemma characterizes the
desirable statistical properties of $\theta$.
\begin{lemma}
\label{stat-theta} Let $\theta$ be as  in
Theorem~\ref{ConvProof1} and $r$ as in
eqn.~(\ref{def_r}). Define the m.s.e. $\zeta$ as
\begin{equation}
\label{mserror} \zeta = \mathbb{E}\left[\left(\mathbf{\theta} -
r\right)^{2}\right]
\end{equation}
Then, the consensus limit~$\theta$ is unbiased and its m.s.e.~bounded as below.
%
\begin{eqnarray}
\label{unb-theta}
\mathbb{E}\left[\theta\right]&=&r\\
%
\label{msbound}
\zeta
&\leq& \frac{\eta}{N^{2}}\sum_{i\geq 0}\alpha^{2}(i)
\end{eqnarray}
%
\end{lemma}
\begin{proof}
From eqn.~(\ref{th2.5}), we have
\begin{equation}
\label{lm:stattheta1}
\mathbb{E}\left[x_{\mbox{\scriptsize{avg}}}(i)\right]=r,\:\:\forall
i
\end{equation}
Since $\left\{x_{\mbox{\scriptsize{avg}}}(i)\right\}_{i\geq 0}$
converges in $\mathcal{L}_{2}$ to $\theta$
(Theorem~\ref{ConvProof1}), it also converges in
$\mathcal{L}_{1}$, and we have
\begin{equation}
\label{lm:stattheta2}
\mathbb{E}\left[\theta\right]=\lim_{i\rightarrow\infty}\mathbb{E}\left[x_{\mbox{\scriptsize{avg}}}(i)\right]=r
\end{equation}
which proves~(\ref{unb-theta}). For~(\ref{msbound}),
by Theorem~\ref{ConvProof1}, the sequence
$\left\{(x_{\mbox{\scriptsize{avg}}}(i)-r)^{2}\right\}_{i\geq 0}$
converges in $\mathcal{L}_{2}$ to $(\theta - r)^{2}$. Hence,
\vspace*{-.5cm}
\begin{eqnarray}
\label{lm:stattheta3-def} \zeta & = & \mathbb{E}\left[\left(\mathbf{\theta}
- r\right)^{2}\right]\nonumber \\ & = &
\lim_{i\rightarrow\infty}\mathbb{E}
\left[\left(x_{\mbox{\scriptsize{avg}}}(i)-r\right)^{2}\right]\nonumber
\\
\label{lm:stattheta3}
& = & \sum_{i\geq
0}\frac{\alpha^{2}(i)}{N^{2}}\mathbb{E}\left[\|\mathbf{n}(i)\|^{2}\right]
\end{eqnarray}
Eqn.~(\ref{msbound}) follows from~(\ref{lm:stattheta3}) and~(\ref{th2.7}), with~$\eta$ the bound on the noise variance.
\end{proof}
Lemma~\ref{stat-thetaGen} in the Appendix shows equivalent results under Assumptions~\textbf{1.2)}, \textbf{2.1)}. Proving under
Assumption~\textbf{2.2)} requires more specific information about
the mixing properties of the Markovian noise sequence, which we do not pursue here.

Eqn.~\ref{lm:stattheta3} gives the exact
representation of the m.s.e. As a byproduct, we obtain the following
corollary for an erasure network with identical link failure
probabilities and i.i.d. channel noise.
\begin{corollary}
\label{cor_erasure} Consider an erasure network with $M$
realizable links, identical link failure probability, $p$,
for each link, and the noise sequence, $\{v_{nl}(i)\}_{1\leq
n,l\leq N,~i\geq 0}$, be of identical variance $\sigma^{2}$. Then
the m.s.e.~is
\begin{equation}
\label{cor_erasure1} \zeta=
\frac{2M\sigma^{2}(1-p)}{N^{2}}\sum_{j\geq 0}\alpha^{2}(j)
\end{equation}
\end{corollary}

\begin{proof}
Using the fact that the link failures and the channel noise are
independent, we have
\begin{equation}
\label{cor_erasure2} \mathbb{E}[\|\mathbf{n}(i)\|^{2}]=
2\sigma^{2}M (1-p)
\end{equation}
The result then follows from eqn.~(\ref{lm:stattheta3}).
\end{proof}
While interpreting the dependence of eqn.~(\ref{cor_erasure1}) on
the number of nodes, $N$, it should be noted that the number of
realizable links $M$ must be $\Omega(N)$ for the connectivity
assumptions to hold.

Lemma~\ref{stat-theta} shows that, for a given bound
on the noise variance, $\eta$, the m.s.e.~$\zeta$ can be made arbitrarily
small by properly scaling the weight sequence, $\{\alpha(j)\}_{j\geq 0}$.
As an example, consider the weight sequence,
\[
\alpha(j)=\frac{1}{j+1},\:\:\forall j
\]
Clearly, this choice of $\alpha(i)$ satisfies the persistence
conditions of eqn.~(\ref{alphacond-1}) and, in fact,
\[
\sum_{j\geq 0}\alpha^{2}(j)=\sum_{j\geq
1}\frac{1}{j^{2}}=\frac{\pi^{2}}{6}
\]
Then, for any $\epsilon>0$, the scaled weight sequence,
$\left\{\widetilde{\alpha}(j)\right\}_{j\geq 0}$,
\[
\widetilde{\alpha}(j)=\frac{\sqrt{6\epsilon}N}{\sqrt{\eta}\pi
(j+1)}
\]
will guarantee that $\zeta\leq\epsilon$. However, reducing the
m.s.e. by scaling the weights in this way will reduce the
convergence rate of the algorithm; this trade-off is considered in
the next Subsection.


\subsection{Convergence Rate}
\label{ConvRateAND} The $\mathcal{A-ND}$ algorithm falls under the
framework of stochastic approximation and, hence, a detailed
convergence rate analysis can be done through the ODE method (see,
for example, \cite{Kushner-Yin}.) For clarity of presentation, we
skip this detailed analysis here; rather, we present a simpler
convergence rate analysis, involving the mean state vector
sequence only under assumptions~\textbf{1.1)}, \textbf{2.1)}, and
\textbf{3)}. From the asymptotic unbiasedness of $\theta$, it
follows
\begin{equation}
\label{cr1}
\lim_{i\rightarrow\infty}\mathbb{E}\left[\mathbf{x}(i)\right]=r\mathbf{1}
\end{equation}
Our goal is to determine the rate at which the sequence
$\{\mathbb{E}\left[\mathbf{x}(i)\right]\}_{i\geq 0}$ converges to
$r\mathbf{1}$.
 Since $L(i)$ and $\mathbf{x}(i)$ are independent, and
$\mathbf{n}(i)$ is zero mean, we have
\begin{equation}
\label{cr2}
\mathbb{E}\left[\mathbf{x}(i+1)\right]=\left(I-\alpha(i)\overline{L}\right)\mathbb{E}\left[\mathbf{x}(i)\right],~\forall
i
\end{equation}
By the persistence condition~(\ref{alphacond}) the sequence
$\alpha(i)\rightarrow 0$. Without loss of generality, we can
assume that
\begin{equation}
\label{cr3}
\alpha(i)\leq\frac{2}{\lambda_{2}\left(\overline{L}\right)+\lambda_{N}(\overline{L})},~\forall
i
\end{equation}
 Then, it can be shown that (see~\cite{tsp07-K-M})
\begin{equation}
\label{cr4}
\left\|\mathbb{E}\left[\mathbf{x}(i)\right]-r\mathbf{1}\right\|\leq
\left( \prod_{0\leq j\leq
i-1}\left(1-\alpha(j)\lambda_{2}\left(\overline{L}\right)\right)\right)\left\|\mathbb{E}\left[\mathbf{x}(0)\right]-r\mathbf{1}\right\|
\end{equation}
Now, since  $1-a\leq e^{-a}$, $0\leq a\leq 1$, we
have
\begin{equation}
\label{cr5}
\left\|\mathbb{E}\left[\mathbf{x}(i)\right]-r\mathbf{1}\right\|\leq\left(e^{-\lambda_{2}\left(\overline{L}\right)\left(\sum_{0\leq
j\leq
i-1}\alpha(j)\right)}\right)\left\|\mathbb{E}\left[\mathbf{x}(0)\right]-r\mathbf{1}\right\|
\end{equation}
Eqn.~(\ref{cr5}) shows that the rate of convergence of the mean consensus depends on the
topology through the algebraic connectivity
$\lambda_2\left(\overline{L}\right)$ of the mean graph and through
the weights $\alpha(i)$. It is interesting to note that the way
the random link failures affect the convergence rate (at least for
the mean state sequence) is through $\lambda_{2}(\overline{L})$, the algebraic connectivity of the \emph{mean} Laplacian, in
the exponent, whereas, for a static network, this reduces to the
algebraic connectivity of the static Laplacian~$L$, recovering the results
in~\cite{Mesbahi-Noise}.

Eqns.~(\ref{cr5}) and~(\ref{msbound})
show a tradeoff between the m.s.e.~and the rate of convergence at
which the sequence
$\left\{\mathbb{E}\left[\mathbf{x}(i)\right]\right\}_{i\geq 0}$
converges to $r\mathbf{1}$. Eqn.~(\ref{cr5}) shows that this rate
of convergence is closely related to the rate at which the weight
sequence, $\alpha(i)$, sums to infinity. For a faster rate, we
want the weights to sum up fast to infinity, i.e., the weights to
be large. In contrast, eqns.~(\ref{msbound}) shows that,
to achieve a small $\zeta$, the weights should be small.

We studied the trade-off between convergence rate and m.s.e.~of
the mean state vectors only.
 In general, more effective
measures of convergence rate are appropriate; intuitively, the
same trade-offs will be exhibited, in the sense that the rate of
convergence will be closely related to the rate at which the
weight sequence, $\alpha(i)$, sums to infinity, as verified by the
numerical studies presented next.

\subsection{Numerical Studies - $\mathcal{A-ND}$}
\label{num_stud_AND} 
\begin{figure}[htb]
\begin{center}
\includegraphics[height=1.8in, width=2in ]{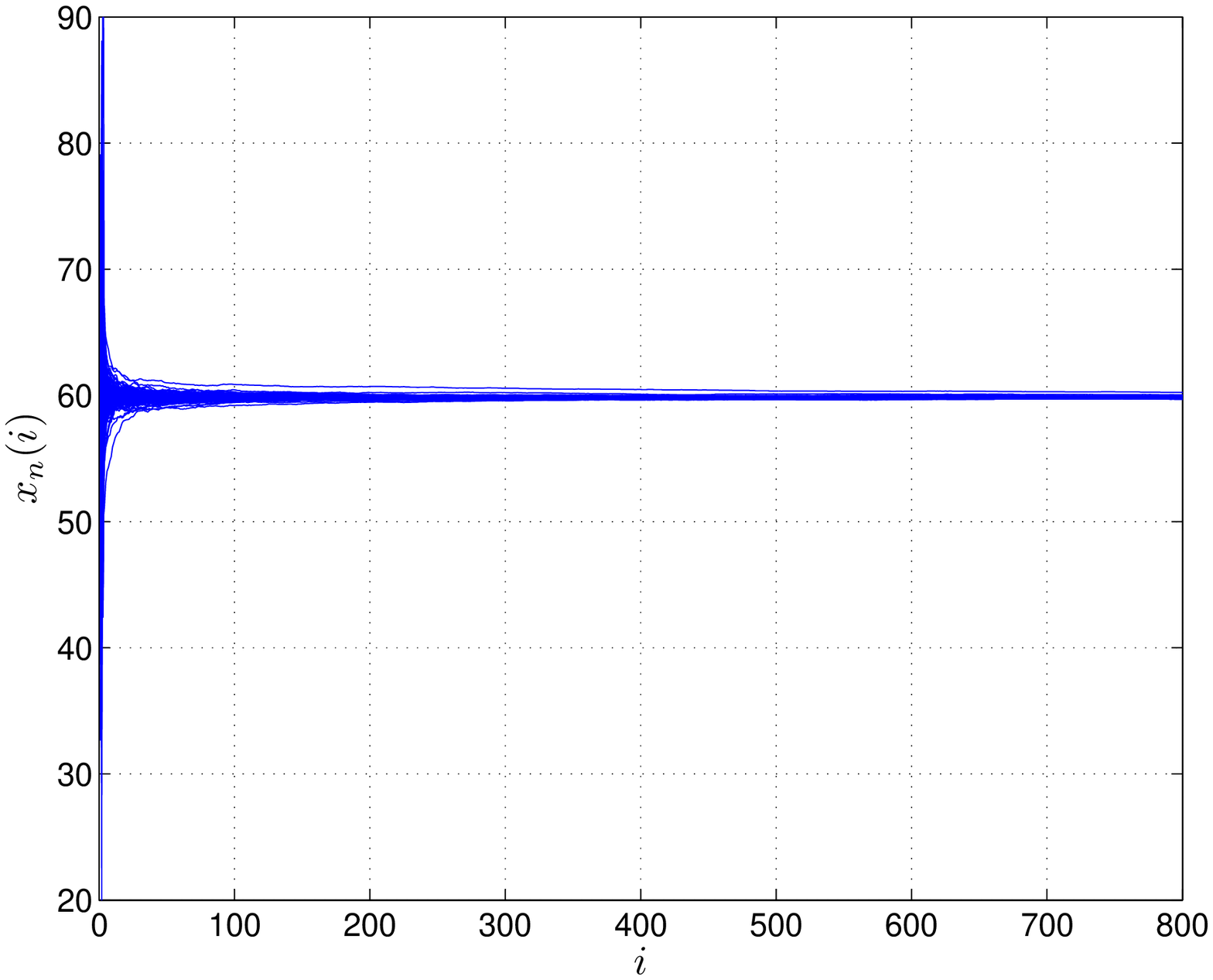}
\includegraphics[height=1.8in, width=2in ]{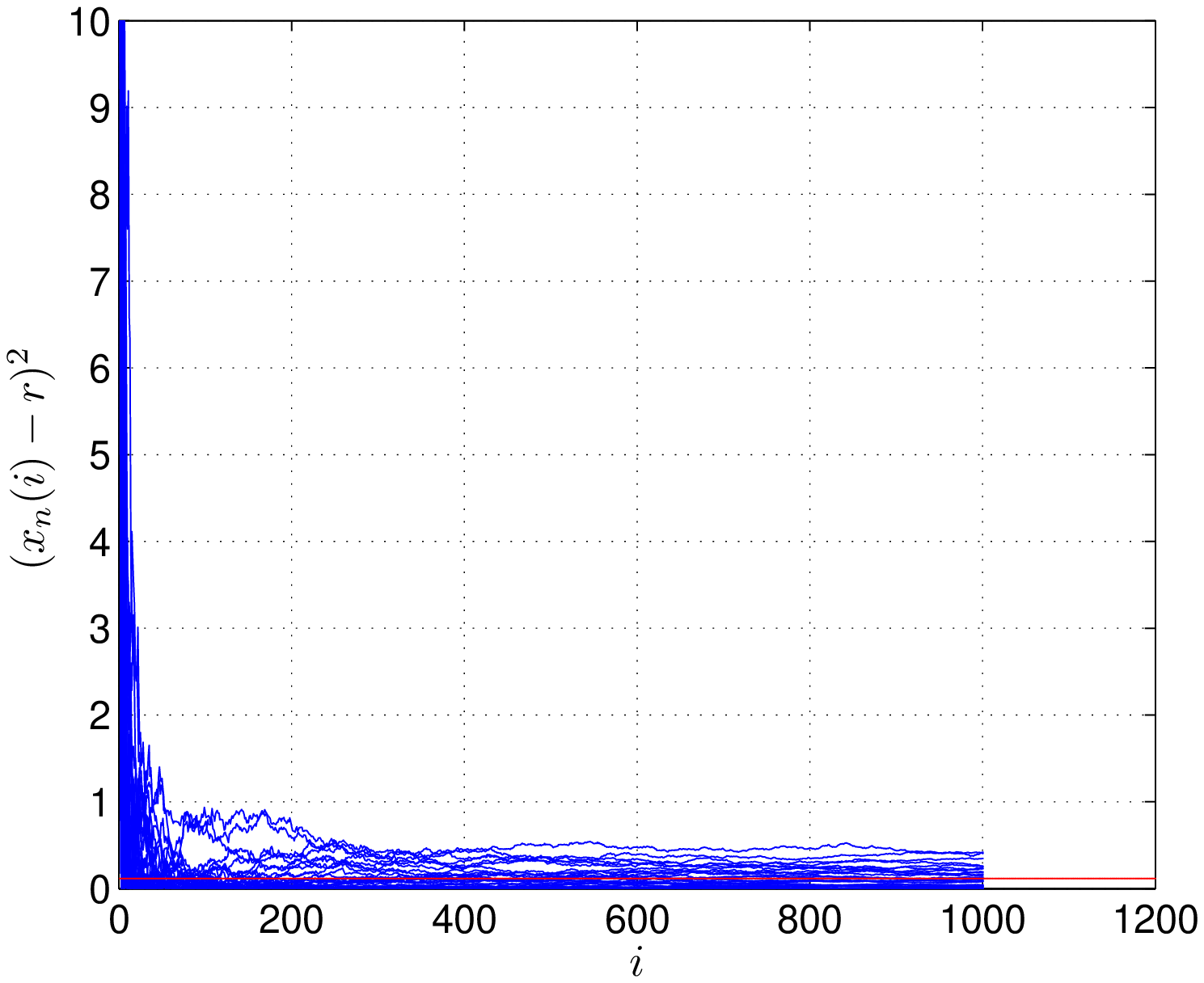}
\includegraphics[height=1.8in, width=2in ]{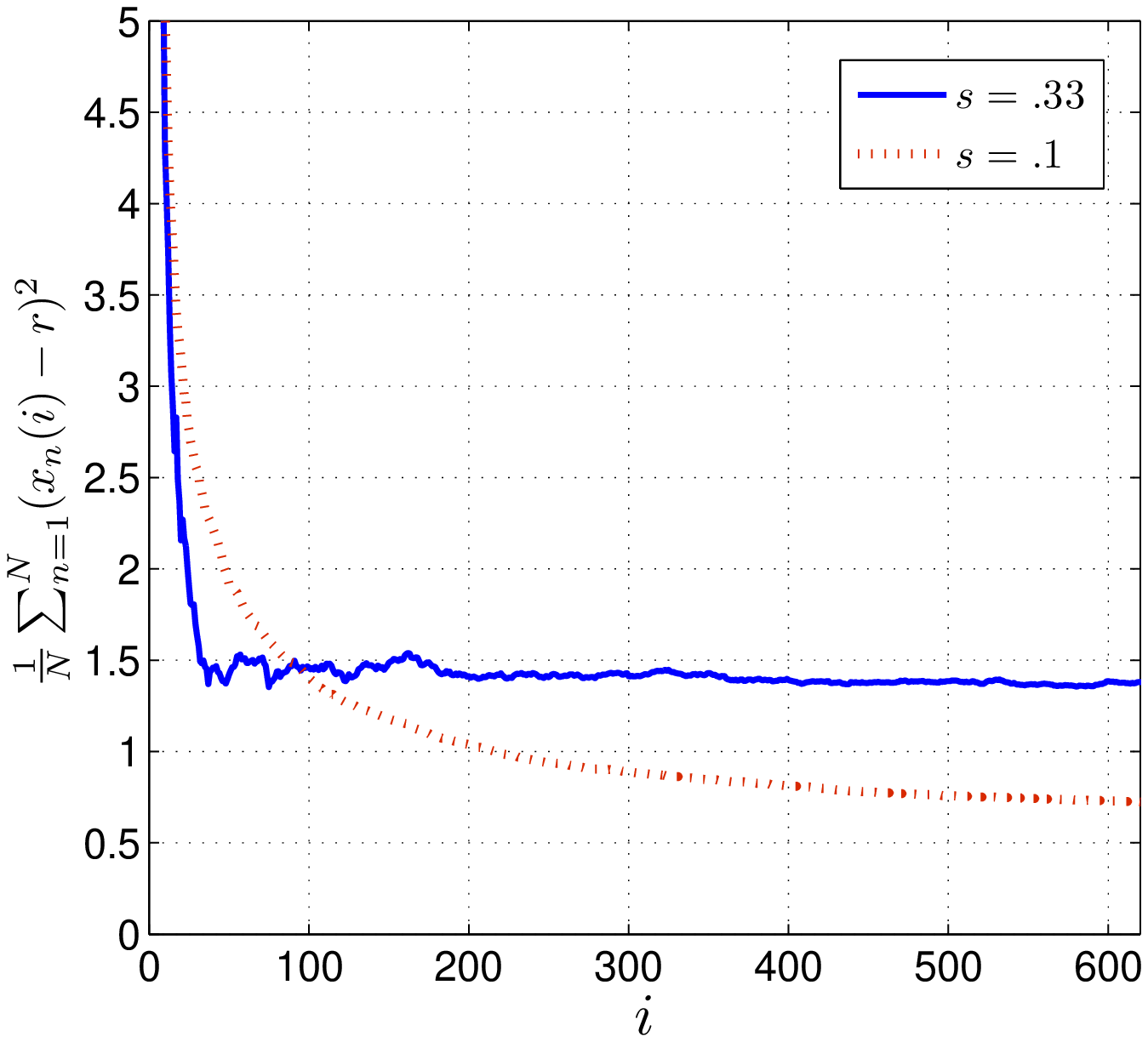}
\caption{Left: Sensors sample paths, $\{x_{n}(i)\}_{1\leq n\leq
N}$, of the $\mathcal{A-ND}$ algorithm, verifying a.s.~consensus.
Center: Sample paths of the $\mathcal{A-ND}$ algorithm, verifying
the m.s.e.~bound. Right: m.s.e~versus convergence rate tradeoff.}
\label{MSE_verify}
\end{center}
\end{figure}

We present numerical studies on the
$\mathcal{A-ND}$ algorithm that verify the analytical results. The
first set of simulations confirms the a.s. consensus in
Theorems~\ref{ConvProof} and~\ref{ConvProof1}. Consider an erasure
network on $N=100$ nodes and $M=5N$ realizable links, with
identical probability of link failure $p=.4$ and identical channel
noise variance $\sigma^{2}=15$. We take $\alpha(i)=1/4i$ and plot
on Fig.~\ref{MSE_verify} on the left, the sample paths
$\{x_{n}(i)\}_{1\leq n\leq N}$, of the sensors over an
instantiation of the $\mathcal{A-ND}$ algorithm. We note that the
sensor states converge to consensus, thus verifying our analytical
results.

The second set of simulations confirms the m.s.e.~in
Corollary~\ref{cor_erasure}. We consider the same erasure network,
but take $\sigma^{2}=30$ and $\alpha(i)=1/5i$. We simulate 50 runs
of the $\mathcal{A-ND}$ algorithm from the initial state.
Fig.~\ref{MSE_verify} on the center plots the propagation of the
squared error $(x_{n}(i)-r)^{2}$ for a randomly chosen sensor $n$
for each of the 50 runs. The cloud of (blue) lines denotes the 50
runs, whereas the extended dashed (red) line denotes the exact
m.s.e. computed in Corollary~\ref{cor_erasure}. The paths are
clustered around the exact m.s.e., thus verifying our results.

The third set of simulations studies the trade-off between m.s.e.
and convergence rate. We consider the same erasure network, but
take $\sigma^{2}=50$, and run the $\mathcal{A-ND}$ algorithm from
the same initial conditions, but for the weight sequences
$\{\alpha_{s}(i)=s/i\}_{i\geq 0}$,  $s = .33, .1$.
Fig.~\ref{MSE_verify} on the right depicts the propagation of the
squared error averaged over all the sensors,
$(1/N)\sum_{n=1}^{N}(x_{n}(i)-r)^{2}$, for each case. We see that
the solid (blue) line ($s = .33$) decays much faster initially
than the dotted (red) line ($s = .1$) and reaches a steady state.
The dotted line ultimately crosses the solid line and continues to
decay at a very slow rate, thus verifying the m.s.e.~versus
convergence rate trade-off, which we established rigorously by
restricting attention to the mean state only.

%
%

\section{$\mathcal{A-NC}$: Consensus with Repeated Averaging}
\label{ConsRepAvg} The stochastic approximation approach to the
average consensus problem, discussed in
Section~\ref{A-NDProbFormAssumptions}, achieves arbitrarily small
m.s.e., see eqn.~(\ref{msbound}), possibly
at the cost of a lower convergence rate, especially, when the
desired m.s.e.~is small. This is mainly because the
weights $\alpha(i)$'s decrease to zero, slowing the convergence as
time progresses, as discussed in Subsection~\ref{ConvRateAND}. In
this Section, we consider an alternative approach based on
repeated averaging, which removes this difficulty. We use a
constant link weight (or step size) and run the consensus
iterations for a fixed number, $\widehat{\imath}$, of iterations.
This procedure is repeated $\widehat{p}$ times, each time starting
from the same initial state $\mathbf{x}(0)$. Since the
$\widehat{p}$ final states obtained at iteration
$\widehat{\imath}$ of each of the $\widehat{p}$ runs are
independent, we average them and get the law of large numbers to
work for us. There is an interesting tradeoff between
$\widehat{\imath}$ and $\widehat{p}$ for a constant total number
of iterations $\widehat{\imath}\widehat{p}$. We describe and
analyze this algorithm and consider its tradeoffs next. The
Section is organized as follows. Subsection~\ref{ProbFormANC} sets
up the problem, states the assumptions, and gives the algorithm
$\mathcal{A-NC}$ for distributed average consensus with noisy
communication links. We analyze the performance of the
$\mathcal{A-NC}$ algorithm in Subsection~\ref{AnalysisConsRepAvg}.
In Subsection~\ref{Numerical-ANC}, we present numerical studies
and suggest generalizations in Subsection~\ref{gen_ANC}.

\subsection{$\mathcal{A-NC}$: Problem Formulation and Assumptions}
\label{ProbFormANC} Again, we consider distributed consensus
with communication channel imperfections in
eqn.~(\ref{genstateupdate}) to
 average the initial state, $\mathbf{x}(0)\in\mathbb{R}^{N\times 1}$.
 The setup is the same as in eqns.~(\ref{def_f_nli}) to~(\ref{A-NDmain1}).

\textbf{$\mathcal{A-NC}$ Algorithm}: The $\mathcal{A-NC}$
algorithm is the following Monte Carlo~(MC) averaging procedure:
\begin{equation}
\label{consrepavg} \mathbf{x}^{p}(i+1)=\mathbf{x}^{p}(i)-\alpha
\left(L\mathbf{x}^{p}(i)+\mathbf{n}^{p}(i)\right),\:\:0\leq i\leq
\widehat{\imath}-1,\:\:1\leq p\leq
\widehat{p},\:\:\mathbf{x}^{p}(0)=\mathbf{x}(0)
\end{equation}
where $x_{n}^{p}(i)$ is the state at sensor~$n$ at the $i$-th
iteration of the $p$-th MC~run. In particular,
$x_{n}^{p}(\widehat{\imath})$ is the state at sensor $n$ at
the end of the $p$-th MC run. Each run of the
$\mathcal{A-NC}$ algorithm proceeds for $\widehat{\imath}$
iterations and there are $\widehat{p}$ MC runs. Finally, the
estimate, $\overline{x}_{n}^{\widehat{p}}(\widehat{\imath})$, of
the average, $x_{\mbox{\scriptsize{avg}}}(0)$, at sensor
$n$ is
\begin{equation}
\label{eq:def_xnp_avg}
\overline{x}_{n}^{\widehat{p}}(\widehat{\imath})=
\frac{1}{\widehat{p}}\sum_{p=1}^{\widehat{p}}x_{n}^{p}(\widehat{\imath})
\end{equation}

We analyze the $\mathcal{A-NC}$ algorithm under the following
assumptions. These make the analysis tractable, but, are not
necessary. They can be substantially relaxed, as shown in
Subsection~\ref{gen_ANC}.

\begin{itemize}[\setlabelwidth{1)}]

\item{\textbf{1)}} \textbf{Static Network}: The Laplacian~$L$ is
fixed (deterministic,) and the network is connected,
$\lambda_{2}(L)>0$. (In Subsection~\ref{gen_ANC} we will allow
random link failures.)
\item{\textbf{2)}} \textbf{Independent Gaussian Noise Sequence}:
The additive noise $\{v^{p}_{nl}(i)\}_{1\leq n,l\leq N,~i,p\geq
0}$ is an independent Gaussian sequence with
\begin{equation}
\label{vnlcond_c}
\mathbb{E}\left[v^{p}_{nl}(i)\right]=0,
\:\:\:\forall 1\leq n,\:l\leq N,\:\:\:i,p\geq
0,\:\:\:
\sup_{n,l,i,p}\mathbb{E}\left[\left(v_{nl}^{p}(i)\right)^{2}\right]=\mu
<\infty
\end{equation}
From eqn.~(\ref{A-NDmain1}), it then follows that
\begin{equation}
\label{ncond_c}\mathbb{E}\left[\mathbf{n}^{p}(i)\right]= \mathbf{0},\:\:\forall i,p\:\:
\sup_{i,l,p}\mathbb{E}\left[\left|n^{p}_{l}(i)\right|^{2}\right]=\phi_{\mbox{\scriptsize{max}}}^{2}\leq
(N-1)\mu <\infty
\end{equation}

\item{\textbf{3)}} \textbf{Constant link weight}:
 The link weight $\alpha$ is constant across
iterations and satisfies
\begin{equation}
\label{alphacond_c}0<\alpha <\frac{2}{\lambda_{N}(L)}
\end{equation}
\end{itemize}
%

%

Let $r=\frac{1}{N}\mathbf{1}^{T}\mathbf{x}(0)$ be the initial
average. To define a uniform convergence metric, assume that the initial
sensor observations $\mathbf{x}(0)$ belong to the following set
(for some $K \in \left[0,\infty\right)$):
\begin{equation}
\label{eq:def_K}
\mathcal{K}=\left\{\mathbf{x}(0)\in\mathbb{R}^{N\times
1}~|~\|\mathbf{x}(0)- r\mathbf{1}\|\leq K\right\}
\end{equation}
 As performance metric for the $\mathcal{A-NC}$ approach,
we adopt the $\epsilon-\delta$ averaging time,
$T^{\alpha}(\epsilon,\delta)$, given by
\begin{equation}
\label{eq:def_Ted} T^{\alpha}(\epsilon,\delta) = i^{\ast} p^{\ast}
\end{equation}
where
\begin{equation}
\label{eq:ed} (i^{\ast},p^{\ast})
=\mbox{arg}~\inf_{\widehat{\imath},\widehat{p}}\left\{(\widehat{\imath}\widehat{p})~|~\inf_{\mathbf{x}(0)\in\mathcal{K}}\inf_{n}\mathbb{P}\left(\frac{|\overline{x}_{n}^{\widehat{p}}(\widehat{\imath})-r|}{K}\leq\epsilon\right)\geq
1-\delta\right\}
\end{equation}
and the superscript $\alpha$ denotes explicitly the dependence on
the link weight $\alpha$.

We say that the $\mathcal{A-NC}$ algorithm achieves
$(\epsilon,\delta)$-\emph{consensus} if
$T^{\alpha}(\epsilon,\delta)$ is finite. A similar notion of
averaging time has been used by others, see for
example,~\cite{Boyd-GossipInfTheory, Shah-SepFunc}. The next
Subsection upper bounds the averaging time and analyzes the
performance of $\mathcal{A-NC}$.
\subsection{Performance Analysis of $\mathcal{A-NC}$}
\label{AnalysisConsRepAvg}
The $\mathcal{A-NC}$ iterations can be rewritten as
\begin{equation}
\label{panc1}
\mathbf{x}^{p}(i+1)=W\mathbf{x}^{p}(i)+\mathbf{\chi}^{p}(i),\:\:0\leq
i\leq \widehat{\imath}-1,\:\:1\leq p\leq
\widehat{p},\:\:\mathbf{x}^{p}(0)=\mathbf{x}(0)
\end{equation}
where
\begin{equation}
\label{panc2} W = I-\alpha
L,\:\:\mathbf{\chi}^{p}(i)=-\alpha\mathbf{n}^{p}(i),~0\leq i\leq
\widehat{\imath}-1,\:\:1\leq p\leq \widehat{p}
\end{equation}
Also, for the choice of $\alpha$ given in
eqn.~(\ref{alphacond_c}), the following can be shown
(see~\cite{Boyd}) for the spectral norm
\begin{equation}
\label{panc3}
\gamma_2=\rho\left(W-\frac{1}{N}J\right)=\rho\left(I-\alpha
L-\frac{1}{N}J\right) < 1, \:\: J=\mathbf{1}\mathbf{1}^T
\end{equation}
where $\rho(\cdot)$ is the spectral radius, which is
equal to the induced matrix 2 norm for symmetric matrices.

We next develop an upper bound on the averaging time
$T^{\alpha}(\epsilon,\delta)$ given in~(\ref{eq:def_Ted}).
Actually, we derive a general bound that holds for generic weight
matrices~$W$. This we do next, in Theorem~\ref{th:RAgen}. We come
back in Theorem~\ref{th:RAmain} to bounding the averaging
time~(\ref{eq:def_Ted}) for the model~(\ref{panc1}) when the
weight matrix~$W$ is as in~(\ref{panc2}) and the spectral norm is
given by~(\ref{panc3}).
\begin{theorem}[Averaging Time]
\label{th:RAgen} Consider the distributed iterative
procedure~(\ref{panc1}). The weight matrix~$W$ is generic, i.e.,
is not necessarily of the form~(\ref{panc2}). It does satisfy the
following assumptions:
\begin{itemize}[\setlabelwidth{1)}]
\item{\textbf{1)}} \textbf{Symmetric Weights}:
\begin{equation}
\label{th:RAgencond1}
W=W^{T},~~W\mathbf{1}=\mathbf{1},~~\gamma_{2}=\rho(W-\frac{1}{N}J)<1
\end{equation}
%
%
%
\item{\textbf{2)}} \textbf{Noise Assumptions}: The sequence
$\{\mathbf{\chi}^{p}(i)\}_{i\geq 0,p\geq 1}$ is a sequence of
independent Gaussian noise vectors with uncorrelated components,
such that
\begin{equation}
\label{th:RAgencond2}
\mathbb{E}\left[\mathbf{\chi}^{p}(i)\right]=\mathbf{0},~~~\sup_{i\geq
0,p\geq 1}\sup_{1\leq l\leq
N}\mathbb{E}\left[(\chi^{p}_{l}(i))^{2}\right]\leq\phi_{\max}^{2}<\infty
\end{equation}
\end{itemize}
Then, we have the following upper bound on the averaging time
$T^{\gamma_{2}}(\epsilon,\delta)$ given by
eqn.~(\ref{eq:def_Ted}):
\begin{eqnarray}
\label{th:RAgen2}
T^{\gamma_{2}}(\epsilon,\delta)&\leq& \widehat{T}^{\gamma_{2}}(\epsilon,\delta)\\
\label{eq:def_appavgtime}
\widehat{T}^{\gamma_{2}}(\epsilon,\delta)&=&
\left(\frac{\ln\epsilon/2}{\ln\gamma_{2}}+1\right)
\left[\left(\frac{4\phi_{\max}^{2}\ln(2/\delta)}{K\epsilon}\right)\left(\frac{\ln\epsilon/2}{N\ln\gamma_{2}}+\frac{1}{N}+
\frac{1-\gamma_{2}^{2}\epsilon^{2}/4}{1-\gamma_{2}^{2}}
\left(1-\frac{1}{N}\right)\right)+1\right]
\end{eqnarray}
(Note we replace the superscript $\alpha$ by $\gamma_{2}$ because
we prove results here for arbitrary $W$ satisfying
eqn.~(\ref{th:RAgencond1}), not necessarily of the form $I-\alpha
L$.)
\end{theorem}

For the proof we need a result from~\cite{tsp06-K-A-M}, which we
state as a Lemma.
\begin{lemma}
\label{lemma-RAgen} Let the assumptions in Theorem~\ref{th:RAgen}
hold true. Define
\begin{equation}
\label{lemma-RAgen1}
m_{l}(\widehat{\imath})=
\mathbb{E}\left[x_{l}^{p}(\widehat{\imath})\right],
\:\:v_{l}(\widehat{\imath})=
\mathbb{E}\left[\left(x_{l}^{p}(\widehat{\imath})
-m_{l}(\widehat{\imath})\right)^{2}\right],\:\:1\leq p\leq
\widehat{p},~~1\leq l\leq N
\end{equation}
(Note that these quantities do not depend on $p$.) Then we have
the following:
\begin{itemize}[\setlabelwidth{1)}]

\item{\textbf{1)}} \textbf{Bound on error mean}:
\begin{equation}
\label{lemma-RAgen2} |m_{l}(\widehat{\imath})
-r|\leq\gamma_{2}^{\widehat{\imath}}\|\mathbf{x}(0)
-r\mathbf{1}\|\leq\gamma_{2}^{\widehat{\imath}}K,
\:\:\forall\mathbf{x}(0)\in\mathcal{K},~~1\leq l\leq N
\end{equation}
\item{\textbf{2)}} \textbf{Bound on error variance}:
\begin{equation}
\label{lemma-RAgen3} v_{l}(\widehat{\imath})\leq
\phi_{\max}^{2}\left[\frac{\widehat{\imath}}{N}+\frac{1-\gamma_{2}^{2\widehat{\imath}}}{1-\gamma_{2}^{2}}(1-\frac{1}{N})\right],
\:\:\forall\mathbf{x}(0)\in\mathbb{R}^{N\times 1},~~1\leq l\leq N
\end{equation}
\item{\textbf{3)}}
$\{x_{l}^{p}(\widehat{\imath})\}_{p=1}^{\widehat{p}}$ is an
i.i.d.~sequence, with
\begin{equation}
\label{lemma-RAgen4}
x_{l}^{p}(\widehat{\imath})\:\eqd\:\mathcal{N}(m_{l}(\widehat{\imath}),v_{l}(\widehat{\imath})),\:\:1\leq
p\leq \widehat{p}
\end{equation}
\end{itemize}
\end{lemma}
\begin{proof}
For proof, see~\cite{tsp06-K-A-M}.
\end{proof}
We now return to the proof of Theorem~\ref{th:RAgen}.

\begin{proof} {[Theorem~\ref{th:RAgen}]} The estimate,
$\overline{x}_{l}^{\widehat{p}}(\widehat{\imath})$, of the
average at each sensor after $\widehat{p}$ runs is in~eqn.~(\ref{eq:def_xnp_avg}). From Lemma~\ref{lemma-RAgen} and a standard Chernoff type bound
for Gaussian random variables (see~\cite{ChungNet}), then
\begin{equation}
\label{th:RAgen5}
\mathbb{P}\left(\frac{\left|\overline{x}_{l}^{\widehat{p}}(\widehat{\imath})
-m_{l}(\widehat{\imath})\right|}{K}\leq \epsilon\right)\geq
1-2\exp\left\{
\frac{{-\widehat{p}K\epsilon}}{2v_{l}(\widehat{\imath})}\right\}
\end{equation}
(For the present derivation we assume Gaussian noise; however, the analysis for other noise models may be done by
using the corresponding large deviation rate functions.)

For arbitrary $\epsilon > 0$, define
\begin{equation}
\label{th:RAgen6}
\widehat{\imath}(\epsilon)=\ceil{\frac{\ln\epsilon}{\ln\gamma_{2}}}
\end{equation}
Also, for arbitrary $\epsilon,\delta>0$, define
\begin{equation}
\label{th:RAgen7}
\widehat{p}(\epsilon,\delta)=\ceil{\frac{2v_{l}(\widehat{\imath})}{K\epsilon}\ln\frac{2}{\delta}}
\end{equation}
Then, we have from eqn.~(\ref{lemma-RAgen2}) in
Lemma~\ref{lemma-RAgen},
\begin{equation}
\label{th:RAgen8} \frac{|m_{l}(\widehat{\imath})-r|}{K}
\leq\epsilon/2,\:\:\:\:\forall~\widehat{\imath}\geq
\widehat{\imath}(\epsilon/2),~~1\leq l\leq N
\end{equation}
Also, we have from eqn.~(\ref{th:RAgen5}),
\begin{equation}
\label{th:RAgen9}
\mathbb{P}\left(\frac{|\overline{x}_{l}^{\widehat{p}}(\widehat{\imath})-m_{l}(\widehat{\imath})|}{K}\leq\epsilon/2\right)\geq
1-\delta,\:\:\:\:\forall\:\:\widehat{p}\geq
\widehat{p}(\epsilon/2,\delta),~~1\leq l\leq N
\end{equation}
From the triangle inequality, we get
\begin{equation}
\label{th:RAgen10}
|\overline{x}_{l}^{\widehat{p}}(\widehat{\imath}) -r|\leq
|\overline{x}_{l}^{\widehat{p}}(\widehat{\imath})
-m_{l}(\widehat{\imath})| +|m_{l}(\widehat{\imath})-r|,~~\forall
1\leq l\leq N
\end{equation}
It then follows from eqn.~(\ref{th:RAgen8}) that
\begin{equation}
\label{th:RAgen11}
|\overline{x}_{l}^{\widehat{p}}(\widehat{\imath})
-m_{l}(\widehat{\imath})|\leq K
\epsilon/2~\Longrightarrow~|\overline{x}_{l}^{\widehat{p}}(\widehat{\imath})
-r|\leq K \epsilon,\:\:\:\forall~\widehat{\imath}\geq
\widehat{\imath}(\epsilon/2),~~1\leq l\leq N
\end{equation}
We thus have for $\widehat{\imath}\geq
\widehat{\imath}(\epsilon/2)$ and $\widehat{p}\geq
\widehat{p}(\epsilon/2,\delta)$
\begin{eqnarray}
\label{th:RAgen12}
\mathbb{P}\left(\frac{|\overline{x}_{l}^{\widehat{p}}(\widehat{\imath})
-r|}{K}\leq\epsilon\right) & \geq &
\mathbb{P}\left(\frac{|\overline{x}_{l}^{\widehat{p}}(\widehat{\imath})
-m_{l}(\widehat{\imath})|}{K}\leq\epsilon/2\right)\\
\nonumber & \geq & 1-\delta,~~1\leq l\leq N
\end{eqnarray}
where the last inequality follows from eqn.~(\ref{th:RAgen9}).

From the definition of $T^{\gamma_{2}}(\epsilon,\delta)$, see
eqn.~(\ref{eq:def_Ted}), we then have
\begin{eqnarray}
\label{th:RAgen13} T^{\gamma_{2}}(\epsilon,\delta) & \leq &
\widehat{\imath}(\epsilon/2)\widehat{p}(\epsilon/2,\delta) \\
\nonumber & \leq &
\left(\frac{\ln\epsilon/2}{\ln\gamma_{2}}+1\right)\left(\frac{4v_{l}(\widehat{\imath})}{K\epsilon}\ln\frac{2}{\delta}+1\right)
\\ \nonumber & \leq &
\left(\frac{\ln\epsilon/2}{\ln\gamma_{2}}+1\right)
\left[\left(\frac{4\phi_{\max}^{2}\ln(2/\delta)}{K\epsilon}\right)\left(\frac{\ln\epsilon/2}{N\ln\gamma_{2}}+\frac{1}{N}+
\frac{1-\gamma_{2}^{2}\epsilon^{2}/4}{1-\gamma_{2}^{2}}
\left(1-\frac{1}{N}\right)\right)+1\right]\nonumber \\ & = &
\widehat{T}^{\gamma_{2}}(\epsilon,\delta)
\end{eqnarray}
where the third inequality follows from eqn.~(\ref{lemma-RAgen3}).
\end{proof}
We call the upper bound on $T^{\gamma_{2}}(\epsilon,\delta)$
in~(\ref{eq:def_appavgtime}) given in Theorem~\ref{th:RAgen}  the
\emph{approximate averaging time}.
We use $\widehat{T}^{\gamma_{2}}(\epsilon,\delta)$ to characterize
the convergence rate of $\mathcal{A-NC}$. We state a property of
$\widehat{T}^{\gamma_{2}}(\epsilon,\delta)$.
\begin{lemma}
\label{lemma-appavgtime} Recall the spectral radius, $\gamma_{2}$,
defined in eqn.~(\ref{th:RAgencond1}). Then, for
$0<\epsilon,\delta<1$, $\widehat{T}^{\gamma_{2}}(\epsilon,\delta)$
is an increasing function of $\gamma_{2}$ in the interval
$0<\gamma_{2}<1$.
\end{lemma}
\begin{proof}
The lemma follows by differentiating
$\widehat{T}^{\gamma_{2}}(\epsilon,\delta)$ with respect to
$\gamma_{2}$.
\end{proof}

We now study the convergence properties of the
$\mathcal{A-NC}$ algorithm, eqn.~(\ref{consrepavg}),
i.e., when the weight matrix~$W$ is of the form~(\ref{panc2}) and
the spectral norm satisfies~(\ref{panc3}). Then,
the averaging time becomes a function of the weight $\alpha$. We make this explicit by denoting as $T^{\alpha}(\epsilon,\delta)$
and $\widehat{T}^{\alpha}(\epsilon,\delta)$ the averaging time and
the approximate averaging time, respectively.
\begin{theorem}[$\mathcal{A-NC}$ Averaging Time]
\label{th:RAmain} Consider the $\mathcal{A-NC}$ algorithm for
distributed averaging, under the assumptions given in
Subsection~\ref{ProbFormANC}. Define
\begin{equation}
\label{re_def} T^{\ast}(\epsilon,\delta) = \inf_{0<\alpha
<\frac{2}{\lambda_{N}(L)}}T^{\alpha}(\epsilon,\delta),~~~\widehat{T}^{\ast}(\epsilon,\delta)
= \inf_{0<\alpha
<\frac{2}{\lambda_{N}(L)}}\widehat{T}^{\alpha}(\epsilon,\delta)
\end{equation}
Then:
\begin{itemize}[\setlabelwidth{1)}]
\item{\textbf{1)}}
\begin{equation}
\label{th:RAmain1} T^{\alpha}(\epsilon,\delta)
<\infty,~~\forall~0<\epsilon,\delta <1,~~0<\alpha
<\frac{2}{\lambda_{N}(L)}
\end{equation}
This essentially means that the $\mathcal{A-NC}$ algorithm is
realizable for $\alpha$ in the interval
$\left(0,\frac{2}{\lambda_{N}(L)}\right)$.
\item{\textbf{2)}} For $0<\alpha<\frac{2}{\lambda_{N}(L)}$ we have
\begin{eqnarray}
\label{th:RAmain2} T^{\alpha}(\epsilon,\delta)&\leq&
\widehat{T}^{\alpha}(\epsilon,\delta)\\
\label{th:RAmain3} \widehat{T}^{\alpha}(\epsilon,\delta) &=&
\left(\frac{\ln\epsilon/2}{\ln\gamma_{2}}+1\right)
\left[\left(\frac{4\alpha^{2}\phi_{\max}^{2}\ln(2/\delta)}{K\epsilon}\right)
\left(\frac{\ln\epsilon/2}{N\ln\gamma_{2}}+\frac{1}{N}+
\frac{1-\gamma_{2}^{2}\epsilon^{2}/4}{1-\gamma_{2}^{2}}
\left(1-\frac{1}{N}\right)\right)\right.\\
\nonumber
&&\hspace{1in}\left.\phantom{\left(\frac{4\alpha^{2}\phi_{\max}^{2}\ln(2/\delta)}{K\epsilon}\right)}
                +1\right]
\end{eqnarray}
and $\gamma_{2} = \rho(I-\alpha L-\frac{1}{N}J)$.
\item{\textbf{3)}} For a given choice of the pair
$0<\epsilon,\delta<1$, the best achievable averaging time
$T^{\ast}(\epsilon,\delta)$ is bounded above by
$\widehat{T}^{\ast}(\epsilon,\delta)$, given by
\begin{eqnarray}
\label{th:RAmain4} \widehat{T}^{\ast}(\epsilon,\delta) & = &
\inf_{0<\alpha\leq\frac{2}{\lambda_{2}(L)+\lambda_{N}(L)}}\left(\frac{\ln\epsilon/2}{\ln(1-\alpha\lambda_{2}(L))}+1\right)
\left[\left(\frac{4\alpha^{2}\phi_{\max}^{2}\ln(2/\delta)}{K\epsilon}\right)\left(\frac{\ln\epsilon/2}{N\ln(1-\alpha\lambda_{2}(L))}\right.\right.
\nonumber
\\ & & \left.\left. +\frac{1}{N}+
\frac{1-(1-\alpha\lambda_{2}(L))^{2}\epsilon^{2}/4}{1-(1-\alpha\lambda_{2}(L))^{2}}
\left(1-\frac{1}{N}\right)\right)+1\right]
\end{eqnarray}
%
\end{itemize}
\end{theorem}
Note that in~(\ref{th:RAmain4}) the optimization is over a smaller
range on~$\alpha$ than in~(\ref{re_def}).

\begin{proof} The iterations for the $\mathcal{A-NC}$ algorithm are given by~(\ref{panc1}), where the weight matrix~$W$ is given by~(\ref{panc2}) and the spectral norm $\gamma_2$ by~(\ref{panc3}).
  Also, since $\mathbf{\chi}^{p}(i) = -\alpha\mathbf{n}^{p}(i)$, we get
 \begin{equation}
\label{th:RAmain8} \sup_{i\geq 0,p\geq 1}\:\:\sup_{1\leq l\leq
N}\mathbb{E}\left[(\chi^{p}_{l}(i))^{2}\right]\leq\alpha^{2}\phi_{\max}^{2}<\infty
\end{equation}
Then, the assumptions~(\ref{th:RAgencond1},\ref{th:RAgencond2}) in
Theorem~\ref{th:RAgen} are satisfied for $\alpha$ in the
range~(\ref{alphacond_c}) and the two items~(\ref{th:RAmain1})
and~(\ref{th:RAmain3}) follow.
%
%
%

To prove item \textbf{3)}, we note that it follows from
\textbf{2)} that
$T^{\ast}(\epsilon,\delta)\leq\widehat{T}^{\ast}(\epsilon,\delta)$,
where
\begin{eqnarray}
\label{th:RAmain9}
\nonumber
 \widehat{T}^{\ast}(\epsilon,\delta) & = &
\inf_{0<\alpha<\frac{2}{\lambda_{N}(L)}}\widehat{T}^{\alpha}(\epsilon,\delta)
\\
\nonumber
&=&
\inf_{0<\alpha<\frac{2}{\lambda_{N}(L)}}
\left(\frac{\ln\epsilon/2}{\ln\gamma_{2}}+1\right)
\times \\
&&
\times
\nonumber
\left[\left(\frac{4\alpha^{2}\phi_{\max}^{2}\ln(2/\delta)}{K\epsilon}\right)\left(\frac{\ln\epsilon/2}{N\ln\gamma_{2}}+\frac{1}{N}+
\frac{1-\gamma_{2}^{2}\epsilon^{2}/4}{1-\gamma_{2}^{2}}
\left(1-\frac{1}{N}\right)\right)+1\right]
\end{eqnarray}
and $\gamma_{2}=\rho\left(I-\alpha L-\frac{1}{N}J\right)$. Now,
consider the functions
\begin{equation}
\label{th:RAmain10}
g(\gamma_{2})=\frac{4\phi_{\max}^{2}\ln(2/\delta)}{K\epsilon}\left(\frac{
\ln\epsilon/2}{\ln\gamma_{2}}+1\right)\left[\frac
{\ln\epsilon/2}{N\ln\gamma_{2}}+\frac{1}{N}+\frac{1-\gamma_{2}^{2}
\epsilon^{2}/4}{1-\gamma_{2}^{2}}\left(1-\frac{1}{N}\right)\right]
\end{equation}
and
\begin{equation}
\label{th:RAmain210}
h(\gamma_{2})=\left(\frac{\ln\epsilon/2}{\ln\gamma_{2}}+1\right)
\end{equation}
with $\gamma_{2}$ as before. Similar to
Lemma~\ref{lemma-appavgtime}, we can show that, $g(\gamma_{2})$
and $h(\gamma_{2})$ are non-decreasing functions of $\gamma_{2}$.
It can be shown that $\gamma_{2}=\rho\left(I-\alpha
L-\frac{1}{N}J\right)$ attains its minimum value at $\alpha =
\alpha^{\bullet}$ (see~\cite{Boyd}), where
\begin{equation}
\label{th:RAmain11}
\alpha^{\bullet}=\frac{2}{\lambda_{2}(L)+\lambda_{N}(L)}
\end{equation}
We thus have
\begin{equation}
\label{th:RAmain12}
g\left(\rho\left(I-\alpha^{\bullet}L-\frac{1}{N}J\right)\right)\leq
g\left(\rho\left(I-\alpha
L-\frac{1}{N}J\right)\right),~~\alpha^{\bullet}\leq\alpha
<\frac{2}{\lambda_{N}(L)}
\end{equation}
and
\begin{equation}
\label{th:RAmain211}
h\left(\rho\left(I-\alpha^{\bullet}L-\frac{1}{N}J\right)\right)\leq
h\left(\rho\left(I-\alpha
L-\frac{1}{N}J\right)\right),~~\alpha^{\bullet}\leq\alpha
<\frac{2}{\lambda_{N}(L)}
\end{equation}
which implies, that, for $\alpha^{\bullet}\leq\alpha
<\frac{2}{\lambda_{N}(L)}$,
\begin{eqnarray}
\label{th:RAmain13} \widehat{T}^{\alpha}(\epsilon,\delta) & = &
\alpha^{2}g\left(\rho\left(I-\alpha L-\frac{1}{N}J\right)\right) + h\left(\rho\left(I-\alpha L-\frac{1}{N}J\right)\right)\\
\nonumber & \geq & (\alpha^{\bullet})^{
2}g\left(\rho\left(I-\alpha^{\bullet}L-\frac{1}{N}J\right)\right) + h\left(\rho\left(I-\alpha^{\bullet} L-\frac{1}{N}J\right)\right)\\
\nonumber & = & \widehat{T}^{\alpha^{\bullet}}(\epsilon,\delta)
\end{eqnarray}
So, there is no need to consider $\alpha>\alpha^{\bullet}$. This
leads to
\begin{eqnarray}
\label{th:RAmain14} \widehat{T}^{\ast}(\epsilon,\delta) & = &
\inf_{0<\alpha<\frac{2}{\lambda_{N}(L)}}\widehat{T}^{\alpha}(\epsilon,\delta)\\
\nonumber & = &
\inf_{0<\alpha\leq\frac{2}{\lambda_{2}(L)+\lambda_{N}(L)}}\widehat{T}^{\alpha}(\epsilon,\delta)
\end{eqnarray}
Also, it can be shown that (see~\cite{Boyd})
\begin{equation}
\label{th:RAmain15} \gamma_{2}=\rho\left(I-\alpha
L-\frac{1}{N}J\right)=1-\alpha\lambda_{2}(L),~~0<\alpha\leq\frac{2}{\lambda_{2}(L)+\lambda_{N}(L)}
\end{equation}
This, together with eqn.~(\ref{th:RAmain14}) proves item
\textbf{3)}.

\end{proof}

We comment on Theorem~\ref{th:RAmain}. For a given connected
network, it gives explicitly the weight~$\alpha$  for which the
$\mathcal{A-NC}$ algorithm is realizable (i.e., the averaging time
is finite for any $(\epsilon,\delta)$ pair.) For these choices of
$\alpha$, it provides an upper bound on the averaging time. The
Theorem also addresses the problem of choosing the  $\alpha$ that
minimizes this upper bound.\footnote{Note, that the minimizer,
$\alpha^{\ast}$ of the upper bound
$\widehat{T}^{\alpha}(\epsilon,\delta)$ does not necessarily
minimize the actual averaging time $T^{\alpha}(\epsilon,\delta)$.
However, as will be demonstrated by simulation studies, the upper
bound is tight and hence $\alpha^{\ast}$ obtained by minimizing
$\widehat{T}^{\alpha}(\epsilon,\delta)$ is a good enough design
criterion.} There is, in general, no closed form solution for this
problem and, as demonstrated by eqn.~(\ref{th:RAmain4}), even if a
minimizer exists, its value depends on the actual choice of the
pair $(\epsilon,\delta)$. In other words, in general, there is no
uniform minimizer of the averaging time.

\subsection{$\mathcal{A-NC}$: Numerical Studies}
\label{Numerical-ANC} 
\begin{figure}[htb]
\begin{center}
\includegraphics[height=1.8in, width=2in ]{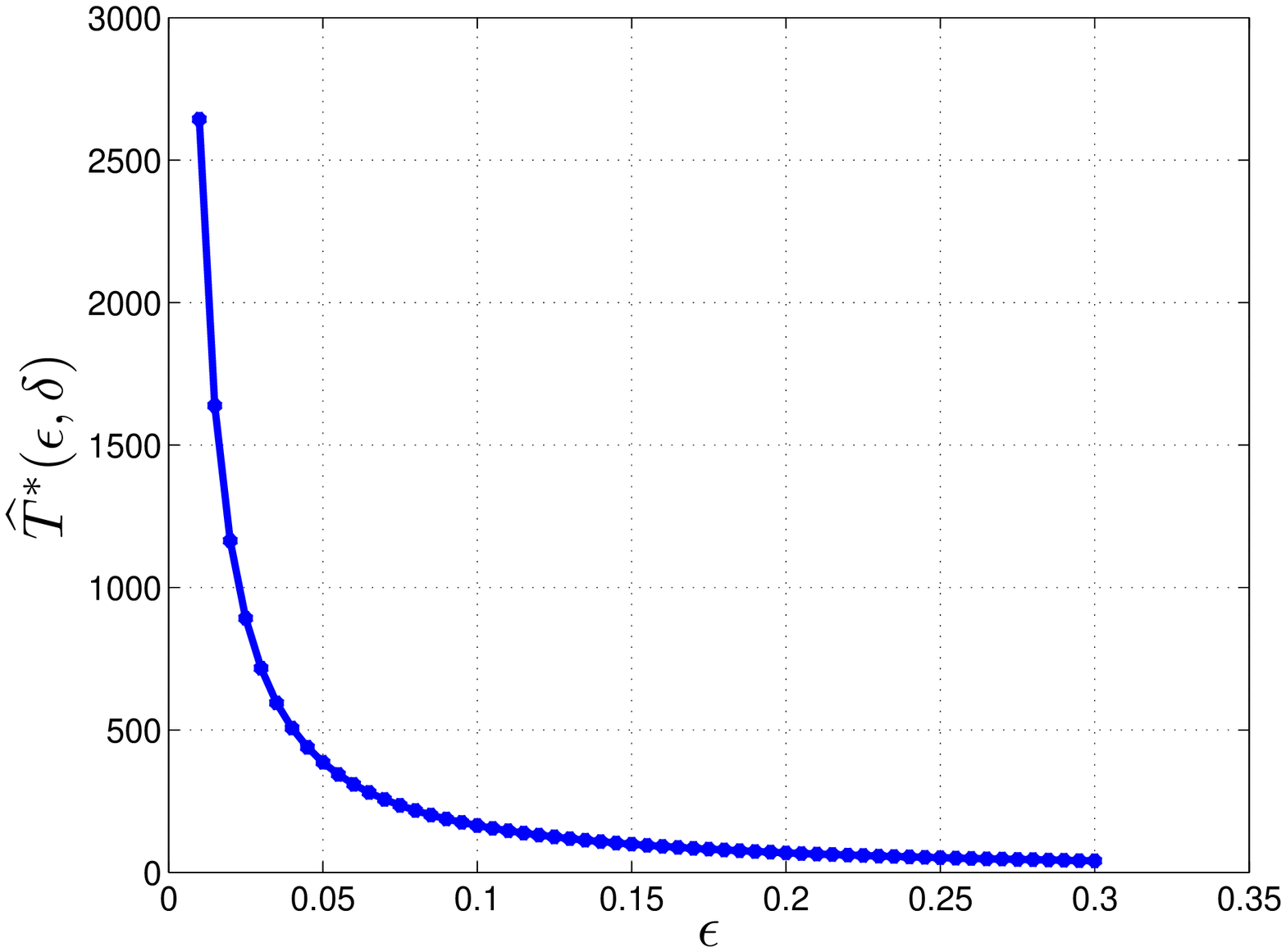}
\includegraphics[height=1.8in, width=2in ]{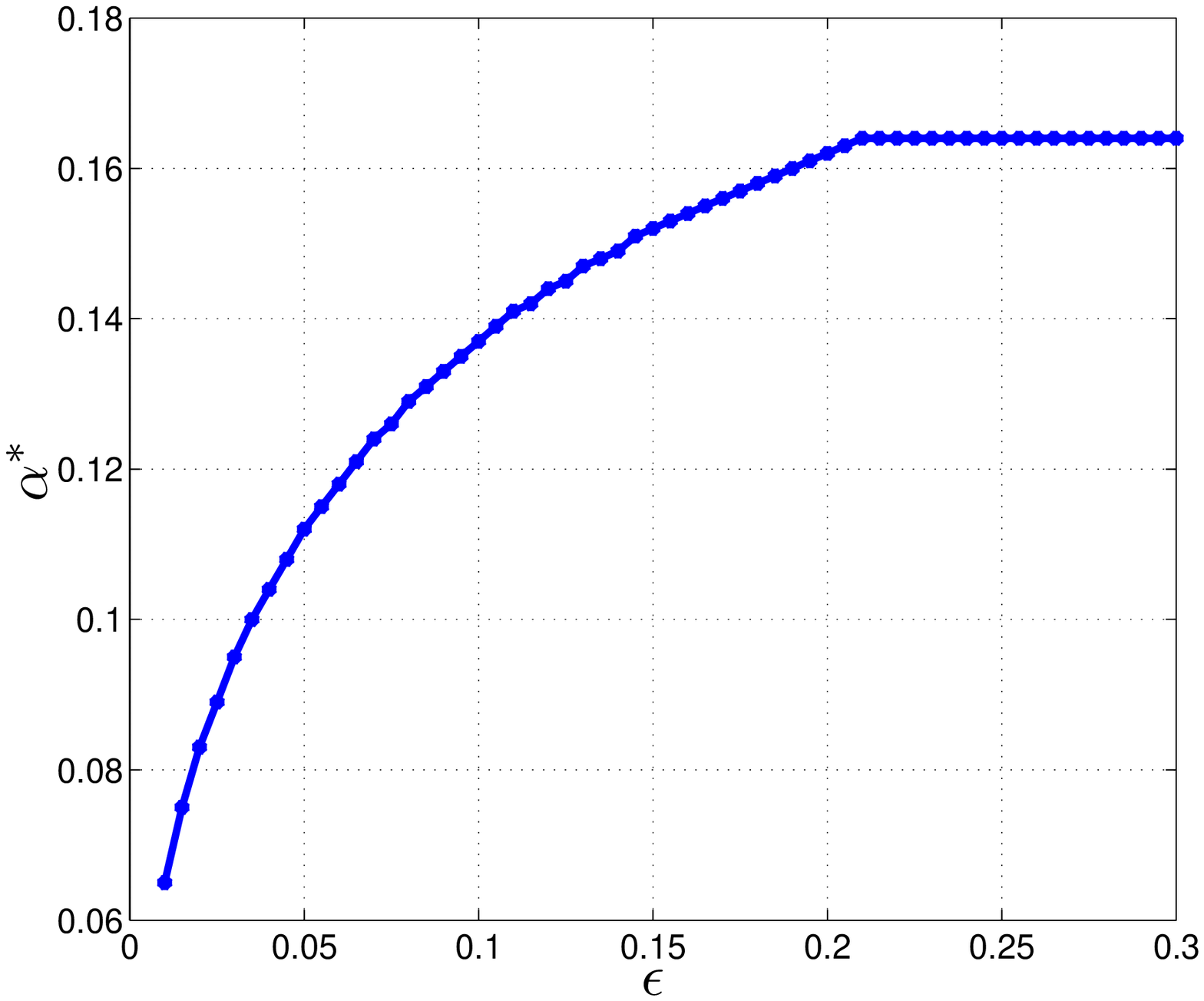}
\caption{Left:Plot of $\widehat{T}^{\ast}(\epsilon,\delta)$ with
varying $\epsilon$, keeping $\delta$ fixed at .05. Right:Plot of
$\alpha^{\ast}$ with varying $\epsilon$, keeping $\delta$ fixed at
.05. } \label{T}
\end{center}
\end{figure}

We present numerical studies on the
$\mathcal{A-NC}$ algorithm. We consider a sensor network of
$N=230$ nodes, with communication topology given by an LPS-II
Ramanujan graph (see~\cite{tsp06-K-A-M}), of degree
6.\footnote{This is a 6-regular graph, i.e., all the nodes have
degree 6.} For the first set of simulations, we take
$\phi^{2}_{\max}=100$, $K=50$, and fix $\delta$ at .05 (this
guarantees that the estimate belongs to the $\epsilon$-ball with
probability at least .95.) We vary $\epsilon$ in steps, keeping
the other parameters fixed, and compute the optimal
$(\epsilon,\delta)$ averaging time,
$\widehat{T}^{\ast}(\epsilon,\delta)$, given by
eqn.~(\ref{th:RAmain4}) and the corresponding optimum
$\alpha^{\ast}$. Fig.~\ref{T} on the left plots
$\widehat{T}^{\ast}(\epsilon,\delta)$ as a function of $\epsilon$,
while Fig.~\ref{T} on the right  plots $\alpha^{\ast}$ vs.
$\epsilon$. As expected, $\widehat{T}^{\ast}(\epsilon,\delta)$
decreases with increasing $\epsilon$. The behavior of
$\alpha^{\ast}$ is interesting. It shows that, to improve accuracy
(small $\epsilon$), the link weight $\alpha$ should be chosen
appropriately small, while, for lower accuracy, $\alpha$ can be
increased, which speeds the algorithm. Also, as $\epsilon$ becomes
larger, $\alpha^{\ast}$ increases to make the averaging time
smaller, ultimately saturating at $\alpha^{\bullet}$, given by
eqn.~(\ref{th:RAmain11}). This behavior is similar to the
$\mathcal{A-ND}$ algorithm, where slower decreasing (smaller)
weight sequences correspond to smaller asymptotic m.s.e.~at a cost
of lower convergence rate (increased accuracy.)

\begin{figure}[htb]
\begin{center}
\includegraphics[height=1.8in, width=2in ]{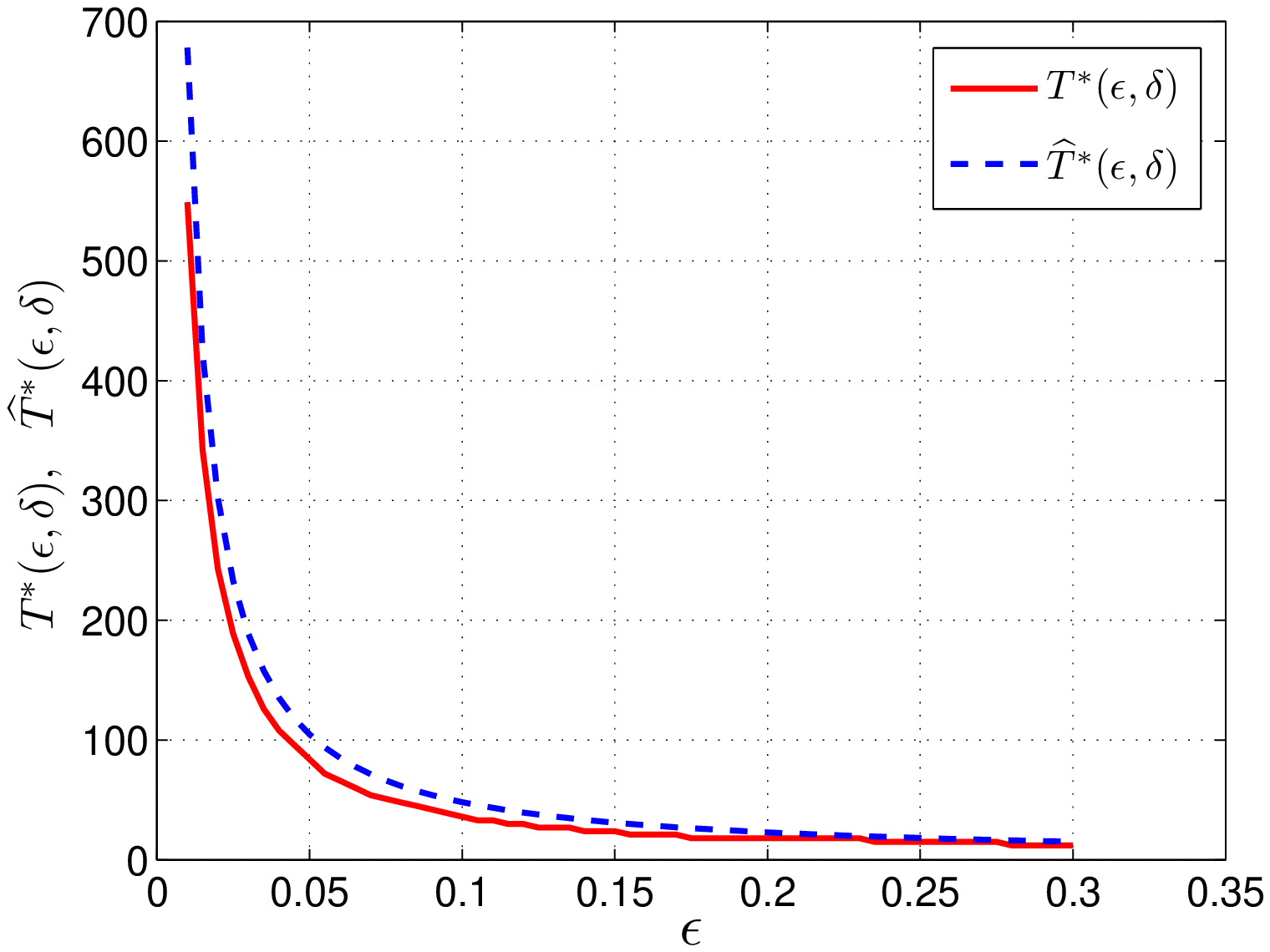}
\includegraphics[height=1.8in, width=2in ]{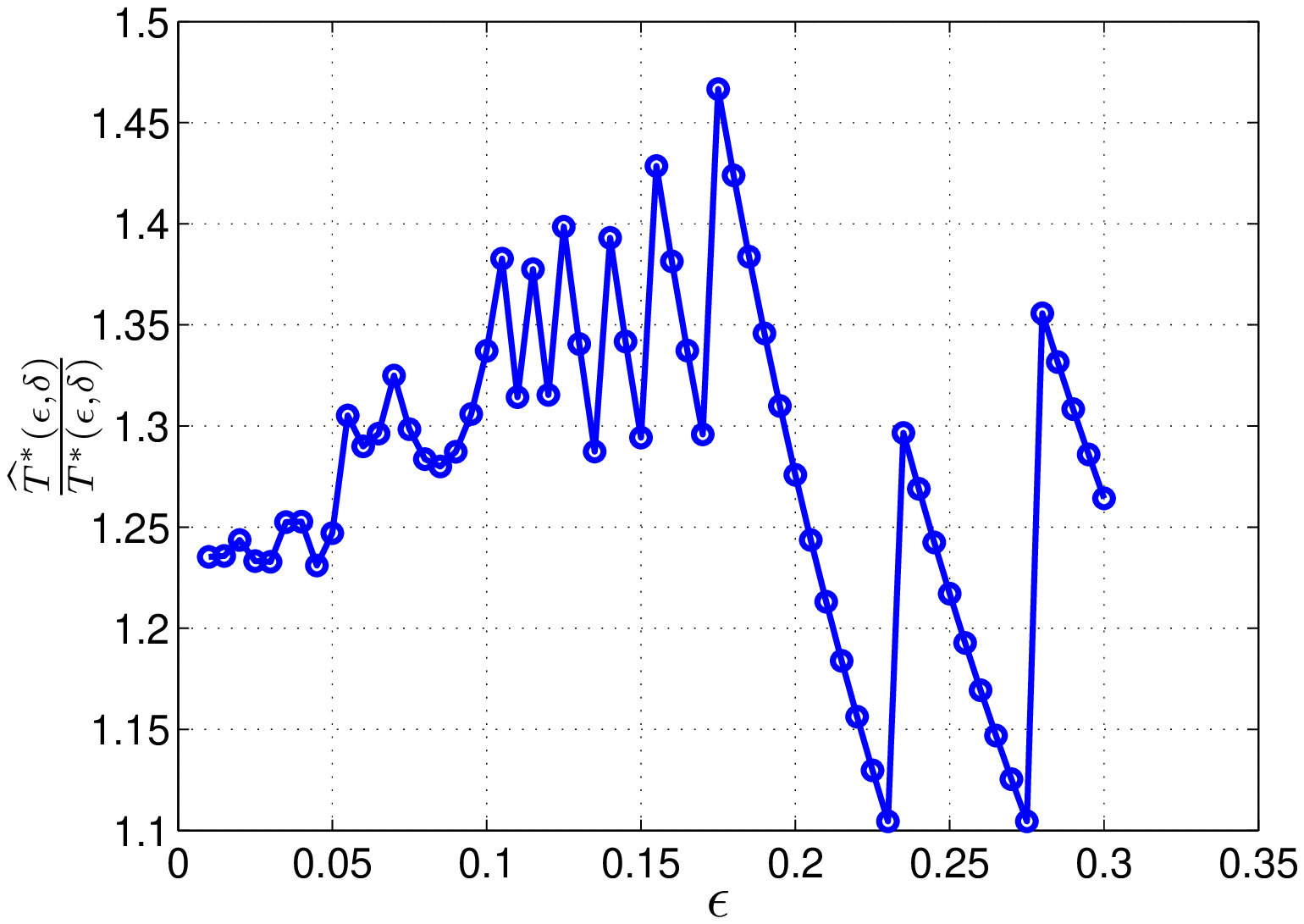}
\caption{Left: Plot of $T^{\ast}(\epsilon,\delta),
\widehat{T}^{\ast}(\epsilon,\delta)$ with varying $\epsilon$.
Right:Plot of
$\frac{\widehat{T}^{\ast}(\epsilon,\delta)}{T^{\ast}(\epsilon,\delta)}$
with varying $\epsilon$. } \label{ANCtight}
\end{center}
\end{figure}

In the second set of simulations, we study the tradeoff between
the number of iterations per Monte-Carlo pass, $\widehat{\imath}$,
and the total number of passes, $\widehat{p}$. Define the quantities, as suggested by
eqn.~(\ref{th:RAgen13}):
\begin{equation}
\label{numstud1} \widehat{i}^{\ast}=
\frac{\ln\epsilon/2}{\ln(1-\alpha^{\ast}\lambda_{2}(L))}+1
\end{equation}
\begin{equation}
\label{numstud2} \widehat{p}^{\ast} = \left(\frac{4\alpha^{\ast
2}\phi_{\max}^{2}\ln(2/\delta)}{K\epsilon}\right)\left(\frac{\widehat{i}^{\ast}}{N}
+ 
\frac{1-(1-\alpha^{\ast}\lambda_{2}(L))^{2}\epsilon^{2}/4}{1-(1-\alpha^{\ast}\lambda_{2}(L))^{2}}
\left(1-\frac{1}{N}\right)\right)+1
\end{equation}
where $\alpha^{\ast}$ is the minimizer in eqn.~(\ref{th:RAmain4}).
In the following, we vary $\epsilon$ and the channel noise
variance $\phi^{2}_{\max}$, taking $K=50$, $\delta = .05$, and
using the same communication network. In particular, in
Fig.~\ref{i,p10} (left) we plot
$(\widehat{i}^{\ast},\widehat{p}^{\ast})$ vs. $\epsilon$ for
$\phi^{2}_{\max}=10$, while in Figs.~\ref{i,p10} (center)
and~\ref{i,p10} (right), we repeat the same for
$\phi^{2}_{\max}=30$ and $\phi^{2}_{\max}=100$ respectively. The
figures demonstrate an interesting trade-off between
$\widehat{i}^{\ast}$ and $\widehat{p}^{\ast}$, and show that for
smaller values of the channel noise variance, the number of
Monte-Carlo passes, $\widehat{p}^{\ast}$ are much smaller than
to the number of iterations per pass, $\widehat{i}^{\ast}$, as
expected. As the channel noise variance grows, i.e., as the
channel becomes more unreliable, we note, that
$\widehat{p}^{\ast}$ increases to combat the noise accumulated at
each pass.
\begin{figure}[htb]
\begin{center}
\includegraphics[height=1.8in, width=2in ]{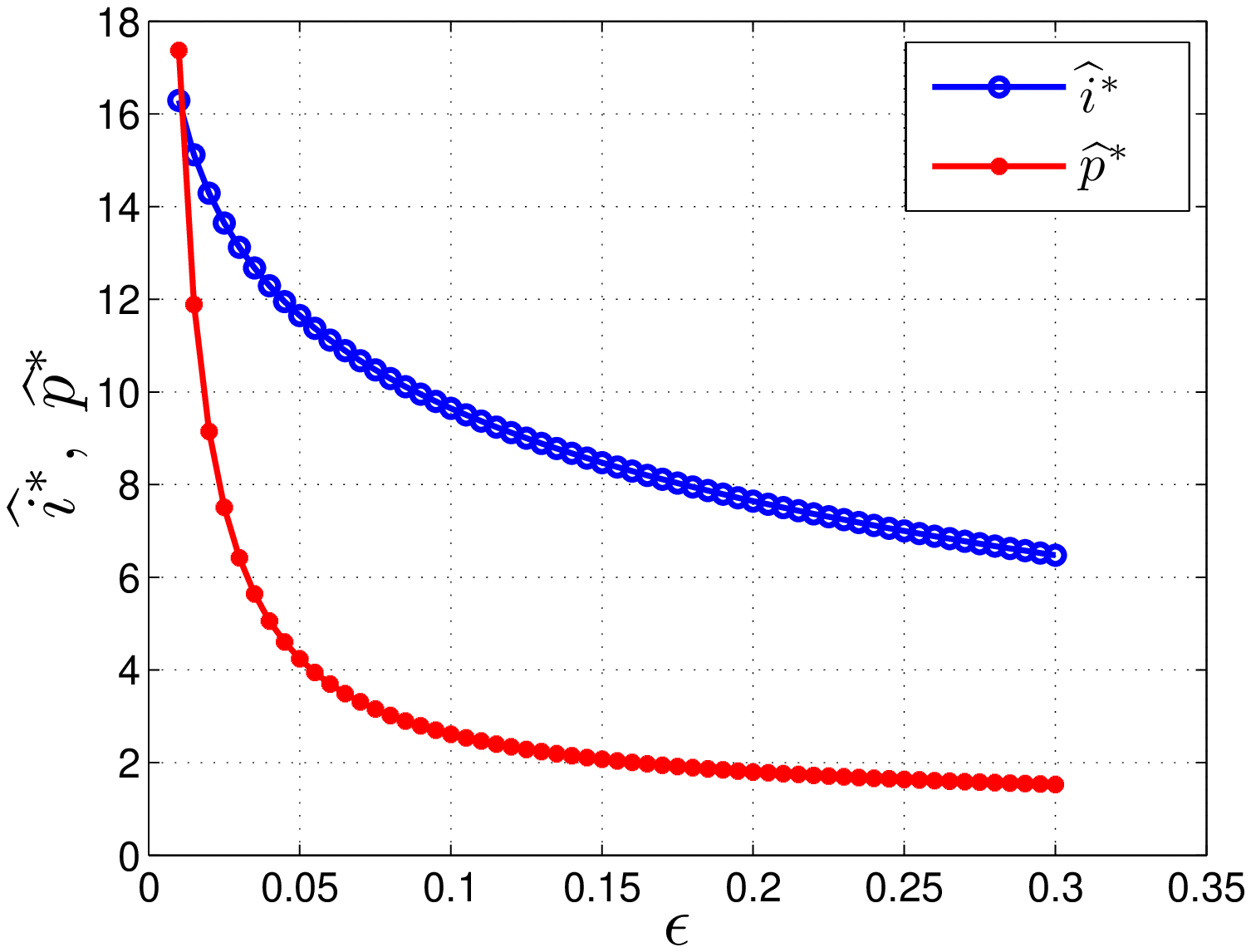}
\includegraphics[height=1.8in, width=2in ]{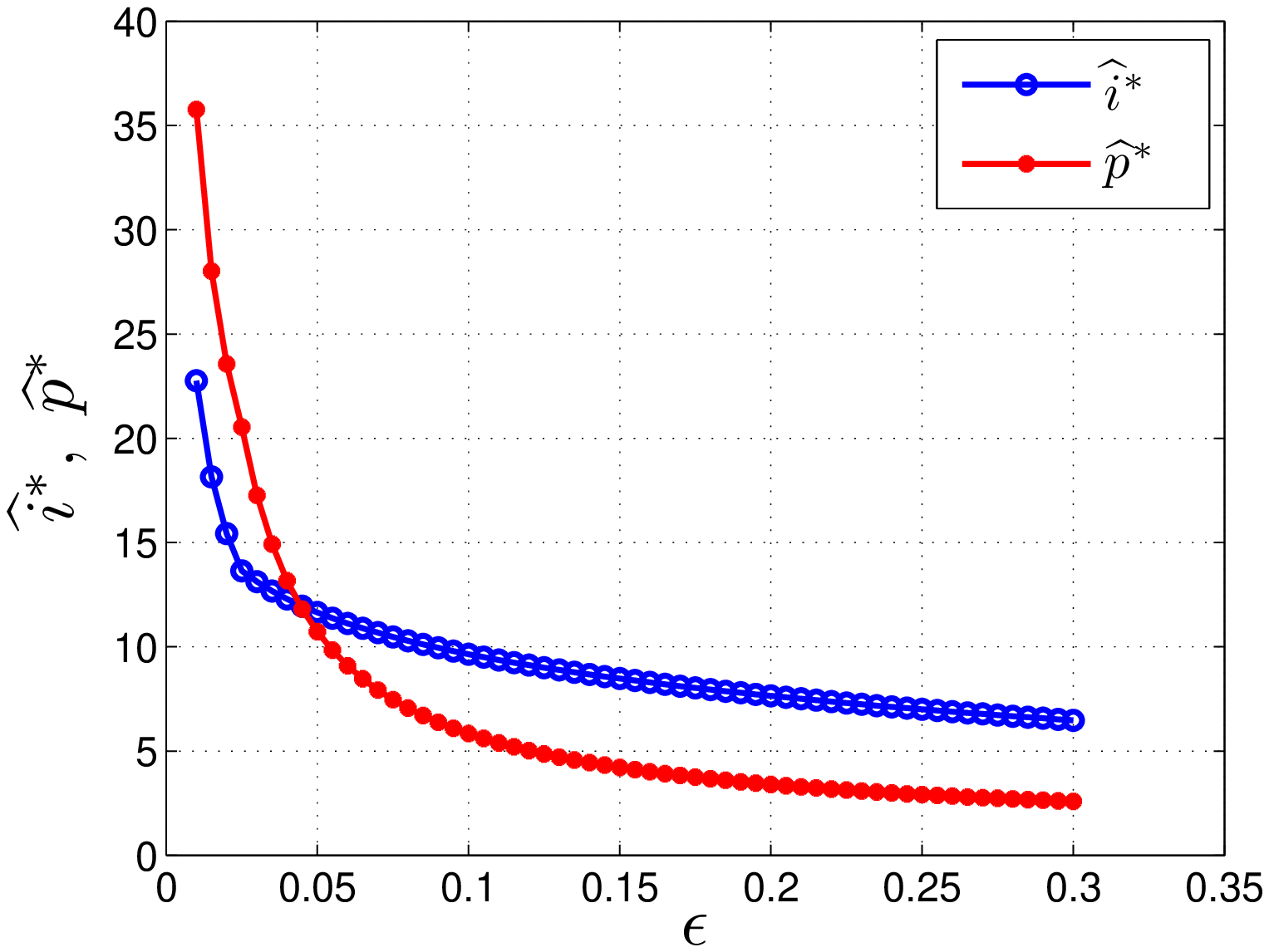}
\includegraphics[height=1.8in, width=2in ]{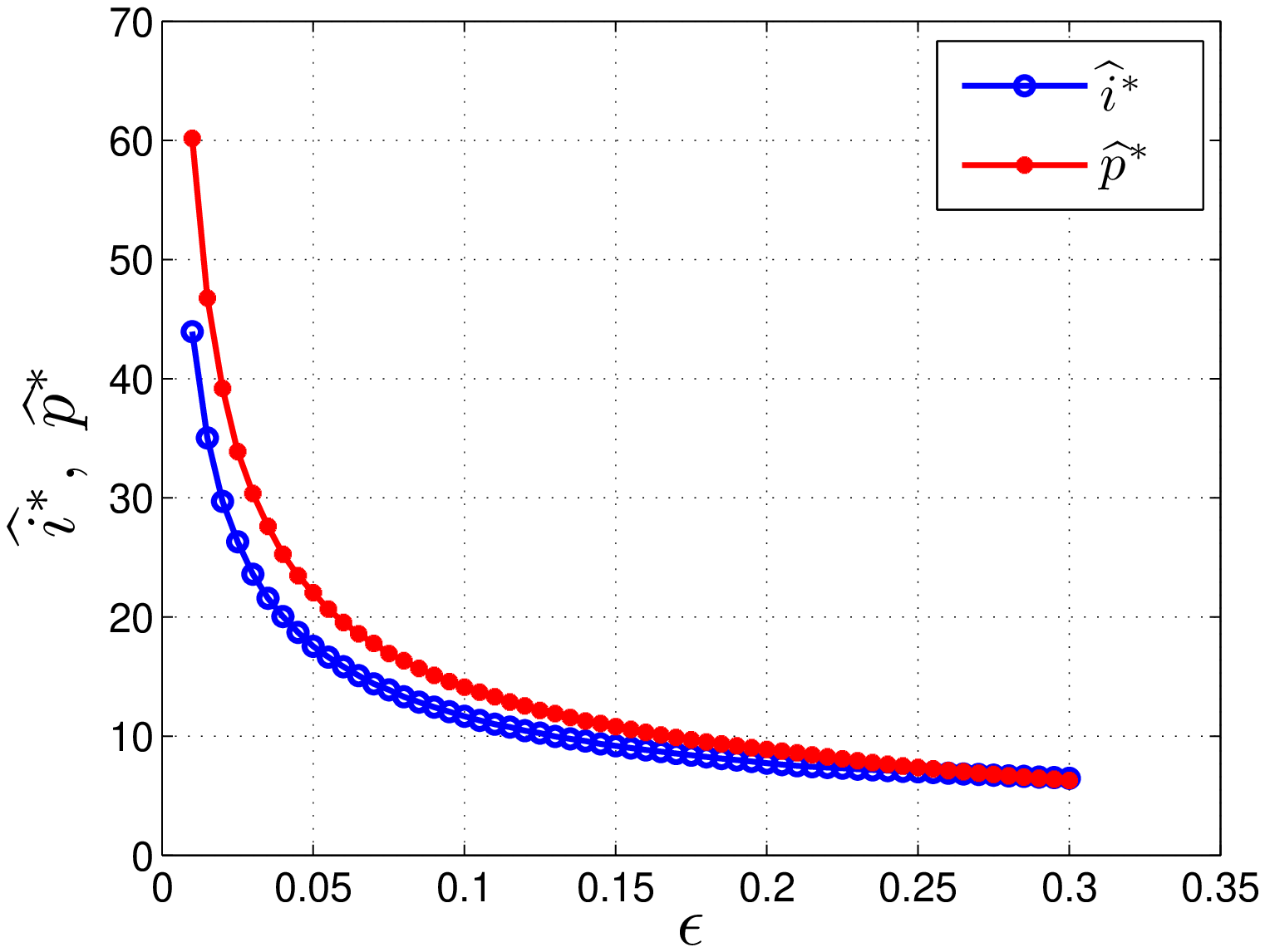}
\caption{Plot of $(\widehat{i}^{\ast},\widehat{p}^{\ast})$ with
varying $\epsilon$: Left:$\phi^{2}_{\max}=10$.
Center:$\phi^{2}_{\max}=30$. Right:$\phi^{2}_{\max}=100$ }
\label{i,p10}
\end{center}
\end{figure}
Finally, we present a numerical study to verify the tightness of
the bound $\widehat{T}^{\ast}(\epsilon,\delta)$. For this we
consider a network of $N=100$ nodes with $M=5N$ edges (generated
according to an Erd\"os-Ren\'{y}i graph, see~\cite{ChungNet}.) We
consider $\phi^{2}_{\max} = 80$, $K=50$, and $\delta=.05$. To
obtain $T^{\ast}(\epsilon,\delta)$ for varying $\epsilon$, we fix
$\alpha$, sample $\mathbf{x}(0)\in\mathcal{K}$ and for each such
$\mathbf{x}(0)$, we generate 100 runs of the $\mathcal{A-NC}$
algorithm. We check the condition in eqn.~(\ref{eq:ed}) and
compute $T^{\alpha}(\epsilon,\delta)$. We repeat this experiment
for varying $\alpha$ to obtain
$T^{\ast}(\epsilon,\delta)=\inf_{\alpha}T^{\alpha}(\epsilon,\delta)$\footnote{The
definition of $T^{\ast}(\epsilon,\delta)$ (see eqn.~(\ref{eq:ed}))
requires the infimum for all $\mathbf{x}(0)\in\mathcal{K}$, which
is uncountable, so we verify eqn.~(\ref{eq:ed}) by sampling points
from $\mathcal{K}$. The $T^{\ast}(\epsilon,\delta)$, thus
obtained, is in fact, a lower bound for the actual
$T^{\ast}(\epsilon,\delta)$.}. We also obtain
$\widehat{T}^{\ast}(\epsilon,\delta)$ from
eqn.~(\ref{th:RAmain4}). Fig.~\ref{ANCtight} on the left
plots $T^{\ast}(\epsilon,\delta)$ (solid red line) and
$\widehat{T}^{\ast}(\epsilon,\delta)$ (dotted blue line) w.r.t.
$\epsilon$, while Fig.~\ref{ANCtight} on the right plots the
ratio $\frac{\widehat{T}^{\ast}(\epsilon,\delta)}{T^{\ast}(\epsilon,\delta)}$
w.r.t. $\epsilon$. The plots show that the bound
$\widehat{T}^{\ast}(\epsilon,\delta)$ is reasonably tight,
especially at small and large values of $\epsilon$, with
fluctuations in between. Note that because
the numerical values obtained for $T^{\ast}(\epsilon,\delta)$  are a lower bound,  the bound
$\widehat{T}^{\ast}(\epsilon,\delta)$ is actually tighter than
appears in the plots.

The bound
$\widehat{T}^{\ast}(\epsilon,\delta)$ is significant since:
 \begin{inparaenum}[i)] \item it is easy to compute from~eqn.~(\ref{th:RAmain4}); \item it is a reasonable approximation to the exact $T^{\ast}(\epsilon,\delta)$;  \item it avoids the costly simulations involved in computing $T^{\ast}(\epsilon,\delta)$ by Monte Carlo; and \item it gives the right tradeoff between $\widehat{i}^{\ast}$ and $\widehat{p}^{\ast}$ (see eqns.~(\ref{numstud1}-\ref{numstud2})), thus
determining the stopping criterion of the $\mathcal{A-NC}$
algorithm.
\end{inparaenum}

\subsection{$\mathcal{A-NC}$: Generalizations}
\label{gen_ANC} In this Subsection, we suggest generalizations to
the $\mathcal{A-NC}$ algorithm. For convenience of analysis, we
assumed before a static network and Gaussian noise. These assumptions can
be considerably weakened. For instance, the static network
assumption may be replaced by a random link failure model with
$\lambda_{2}\left(\overline{L}\right)>0$, where
$\overline{L}=\mathbb{E}\left[L\right]$. Also, the independent
noise sequence in eqn.~(\ref{vnlcond_c}) may be non-Gaussian. In
this case, the $\mathcal{A-NC}$ algorithm iterations will take the
form
\begin{equation}
\label{gen_ANC1}
\mathbf{x}^{p}(i+1)=\left(\mathbf{x}^{p}(i)-\alpha
L^{p}(i)\mathbf{x}^{p}(i)\right)-\alpha\mathbf{n}^{p}(i),~~0\leq
i\leq \widehat{\imath}-1,~1\leq p\leq
\widehat{p},~\mathbf{x}^{p}(0)=\mathbf{x}(0)
\end{equation}
where, $L^{p}(i)$ is an i.i.d.~sequence of Laplacian matrices with
$\lambda_{2}\left(\overline{L}\right)>0$. We then have,
\begin{equation}
\label{gen_ANC2}
\mathbf{x}^{p}(\widehat{\imath})=\left(\prod_{j=0}^{\widehat{\imath}-1}\left(I-\alpha
L^{p}(j)\right)\right)\mathbf{x}(0)+H^{p}(\widehat{\imath}),~1\leq
p\leq \widehat{p}
\end{equation}
It is clear that
$\left\{H^{p}\left(\widehat{\imath}\right)\right\}_{1\leq p\leq
\widehat{p}}$ is an i.i.d. sequence of zero mean random variables.
Under the assumption $\lambda_{2}\left(\overline{L}\right)>0$,
there exists $\alpha$ (see~\cite{tsp07-K-M}) such that
\begin{equation}
\label{gen_ANC4} \mathbb{P}\left[\left(\prod_{j\geq
0}\left(I-\alpha
L^{p}(j)\right)\right)\mathbf{x}(0)=r\mathbf{1}\right]=1,\:\:1\leq
p\leq \widehat{p}
\end{equation}
We thus choose $\widehat{\imath}$ so that
$\left(\prod_{j=0}^{\widehat{\imath}-1}\left(I-\alpha
L^{p}(j)\right)\right)\mathbf{x}(0)$ is sufficiently \emph{close}
to $r\mathbf{1}$. The final estimate is
\begin{equation}
\label{gen_ANC3}
\overline{\mathbf{x}}^{\widehat{p}}\left(\widehat{\imath}\right)=\frac{1}{\widehat{p}}\sum_{p=1}^{\widehat{p}}
\left(\prod_{j=0}^{\widehat{\imath}-1}\left(I-\alpha
L^{p}(j)\right)\right)\mathbf{x}(0)+\frac{1}{\widehat{p}}\sum_{p=1}^{\widehat{p}}H^{p}(\widehat{\imath})
\end{equation}
The first sum is close to $r\mathbf{1}$ by choice of
$\widehat{\imath}$. We now choose $\widehat{p}$, large enough, so
that the second term is \emph{close} to zero. In this way, we can
apply the $\mathcal{A-NC}$ algorithm to more general scenarios.

The above argument guarantees that
$(\epsilon,\delta)$-\emph{consensus} is achievable under these
generic conditions, in the sense that the corresponding averaging
time, $T^{\alpha}(\epsilon,\delta)$, will be finite. A
thorough analysis requires a reasonable computable upper
bound like eqn.~(\ref{th:RAmain3}), followed by  optimization
over $\alpha$ to give the best achievable convergence rate. (Note
a computable upper bound is required, because, as pointed earlier,
it is very difficult to find stopping criterion using Monte-Carlo
simulations, and the resulting
$T^{\ast}(\epsilon,\delta)$ will be a lower bound since
the set $\mathcal{K}$ is uncountable.) One way to proceed is to
identify the appropriate large deviation rate (as suggested in
eqn.~(\ref{th:RAgen5}).) However, the results will depend on the
specific nature of the link failures and noise, which we do not
pursue in this paper due to lack of space.
%

\section{Conclusion}
\label{conclusion} We consider distributed average
consensus when the topology is random (links may fail at random
times) \emph{and} the communication in the channels is corrupted
by additive noise.  Noisy consensus leads to  a bias-variance dilemma.
We considered two versions of consensus that lead to two compromises to this problem:
\begin{inparaenum}[i)] \item
$\mathcal{A-ND}$ fits the framework of stochastic
approximation. It a.s.~converges to the consensus subspace and to a consensus random variable~$\theta$--an unbiased estimate of the desired
average, whose variance we compute and bound;
and \item $\mathcal{A-NC}$ uses repeated averaging
by Monte Carlo, achieving $(\epsilon,\delta)$-\emph{consensus}.
\end{inparaenum}
In $\mathcal{A-ND}$ the bias can be made arbitrarily small, but the rate at which
it decreases can be traded for variance -- trade-off between m.s.e.~and
convergence rate. $\mathcal{A-NC}$ uses a
constant weight $\alpha$ and hence outperforms $\mathcal{A-ND}$ in
terms of convergence rate. Computation-wise,
$\mathcal{A-ND}$ is superior since $\mathcal{A-NC}$  requires more inter-sensor
coordination to execute the independent passes. The estimate
obtained by $\mathcal{A-NC}$ does not possess the
nice statistical properties, including
 unbiasedness, as the computation is terminated after a
finite time in each pass.

Finally,  these algorithms  may be
applied to other problems  in sensor
networks with random links and noise, e.g.,
distributed load balancing in parallel processing or
distributed network flow.

\appendix[Proof of Theorem~\ref{MarkovProc} and $\mathcal{A-ND}$ generalizations under Assumptions~\textbf{1.2)} and \textbf{2.2)}]
\begin{proof}{[Theorem~\ref{MarkovProc}]}
The proof follows that of Theorem~2.7.1
in~\cite{Nevelson}. Suffices to prove it  for
$\mathbf{x}(0)=\mathbf{x}_{0}\:\:\mbox{a.s.}$,
 $\mathbf{x}_{0}\in\mathbb{R}^{N\times 1}$ is a deterministic
starting state. Let the filtration
$\left\{\mathcal{F}_i^{\mathbf{X}}
=\sigma\left\{\mathbf{x}(j): 0\leq j\leq i\right\}\right\}_{i\geq0}$
w.r.t.~which $\left\{\mathbf{x}(i)\right\}_{i\geq0}$,
$\mathbf{x}\in\mathbb{R}^{N\times1}$
$\left(\mbox{and hence functionals of
$\left\{\mathbf{x}(i)\right\}_{i\geq0}$}\right)$ are adapted.

Define the function $W(i,\mathbf{x}),i\geq
0,~\mathbf{x}\in\mathbb{R}^{N\times 1}$ as
\begin{equation}
\label{Wdef}
W(i,\mathbf{x})=\left(1+V\left(i,\mathbf{x}\right)\right)\prod_{j\geq
i}\left[1+g(j)\right]
\end{equation}
It can be shown that
\begin{equation}
\label{Wcond} \mathcal{L}W(i,\mathbf{x})\leq
-\alpha(i)\varphi(i,\mathbf{x}),\:\:i\geq
0,\:\mathbf{x}\in\mathbb{R}^{N\times 1}
\end{equation}
and, hence, under the assumptions (Theorem 2.5.1
in~\cite{Nevelson})
\begin{equation}
\label{rhoinf} \mathbb{P}\left(\liminf_{i\rightarrow\infty}\rho
(\mathbf{x}(i),B)=0\right)=1
\end{equation}
which, together with assumption~(\ref{Vcond3}), implies
\begin{equation}
\label{convmarkov14} \mathbb{P}\left(\liminf_{i\rightarrow\infty}V
\left(i,\mathbf{x}(i)\right)=0\right)=1
\end{equation}
Also, it can be shown that the process
$\left(W(i,\mathbf{x}(i)),\mathcal{F}^{\mathbf{X}}_{i}\right)$ is
a non-negative supermartingale (Theorem~2.2.2 in~\cite{Nevelson})
and, hence, converges a.s. to a finite value. It then follows from
eqn.~(\ref{Wdef}) that $V(i,\mathbf{x}(i))$ also converges a.s. to
a finite value. Together with eqn.~(\ref{convmarkov14}), the a.s.~convergence of $V(i,\mathbf{x}(i))$ implies
\begin{equation}
\label{Vconv}
\mathbb{P}\left(\lim_{i\rightarrow\infty}V(i,\mathbf{x}(i))=0\right)=1
\end{equation}
The theorem then follows from assumptions~(\ref{Vcond1})
and~(\ref{Vcond2}) (see also Theorem~2.7.1 in~\cite{Nevelson}.)
\end{proof}

\begin{theorem}[$\mathcal{A-ND}$: Convergence]
\label{ConvProofgen} Consider the $\mathcal{A-ND}$  algorithm given in
Section~\ref{ProbFormStochApp} with arbitrary initial state
$\mathbf{x}(0)\in\mathbb{R}^{N\times 1}$, under the
Assumptions~\textbf{1.2), 2.2), 3)}. Then,
\begin{equation}
\label{App0}
\mathbb{P}\left[\lim_{i\rightarrow\infty}\rho(\mathbf{x}(i),\mathcal{C})=0\right]=1
\end{equation}
\end{theorem}
\begin{proof}
In the  $\mathcal{A-ND}$ eqn.~(\ref{A-NDmain}),  the Laplacian $L(i,\mathbf{x}(i))$ and the noise $\mathbf{n}(i, \mathbf{x}(i))$ are both state dependent.
 We follow Theorem~\ref{ConvProof} till eqn.~(\ref{assV3}) and
modify eqn.~(\ref{assV3.0}) according to the new assumptions. The sequences $\{L(i,\mathbf{x})\}$ and
$\{\mathbf{n}(i,\mathbf{x})\}$ are independent. By the Gershgorin
circle theorem, the eigenvalues of
$(L(i,\mathbf{x})-\overline{L})$ are less than $2N$ in magnitude.
From the noise variance growth condition we have,
\begin{eqnarray}
\label{App1}
\mathbb{E}\left[\alpha^{2}(i)\mathbf{n}^{T}(i,\mathbf{x})
\overline{L}\mathbf{n}(i,\mathbf{x})\right]
& = &
\alpha^{2}(i)\mathbb{E}\left[\mathbf{n}^{T}_{\mathcal{C}^{\perp}}
\left(i,\mathbf{x}\right)\overline{L}
\mathbf{n}_{\mathcal{C}^{\perp}}\left(i,\mathbf{x}\right)\right]\nonumber
\\ & \leq &
\alpha^{2}(i)\lambda_{N}(L)\mathbb{E}\left[\left\|\mathbf{n}_{\mathcal{C}^{\perp}}
\left(i,\mathbf{x}\right)\right\|^{2}\right]\nonumber
\\ & \leq &
\alpha^{2}(i)\lambda_{N}(L)\left[c_{1}+c_{2}
\left\|\mathbf{x}_{\mathcal{C}^{\perp}}\right\|^{2}\right]
\end{eqnarray}
Using eqn.~(\ref{App1}) and a sequence of steps similar to
eqn.~(\ref{assV3.0}) we have
\begin{eqnarray}
\nonumber
\mathcal{L}V\left(i,\mathbf{x}\right) & = &
-2\alpha(i)\mathbf{x}^{T}\overline{L}^{2}\mathbf{x}+
\alpha^{2}(i)\mathbf{x}^{T}\overline{L}^{3}\mathbf{x}\\
\nonumber
&+&\mathbb{E}\left[\alpha^{2}(i)\left(\left(L\left(i,\mathbf{x}\right)-
\overline{L}\right)\mathbf{x}\right)^{T}\overline{L}
\left(\left(L\left(i,\mathbf{x}\right)-\overline{L}\right)\mathbf{x}\right)\right]
+\mathbb{E}\left[\alpha^{2}(i)\mathbf{n}
\left(i,\mathbf{x}\right)^{T}\overline{L}\mathbf{n}\left(i,\mathbf{x}\right)\right]
\\
\nonumber
& \leq &
-2\alpha(i)\mathbf{x}^{T}\overline{L}^{2}\mathbf{x}
+\alpha^{2}(i)\lambda_{N}\left(\overline{L}\right)^{3}
\left\|\mathbf{x}_{\mathcal{C}^{\perp}}\right\|^{2}
+4\alpha^{2}(i)N^{2}\lambda_{N}\left(\overline{L}\right)
\left\|\mathbf{x}_{\mathcal{C}^{\perp}}\right\|^{2}\\
\nonumber
&&\hspace{6.3cm}+\alpha^{2}(i)\lambda_{N}\left(\overline{L}\right)
\left[c_{1}+c_{2}\left\|\mathbf{x}_{\mathcal{C}^{\perp}}\right\|^{2}\right]
\end{eqnarray}
Now, using the fact that
$\mathbf{x}^{T}\overline{L}\mathbf{x}\geq\lambda_{2}\left(\overline{L}\right)
\left\|\mathbf{x}_{\mathcal{C}^{\perp}}\right\|^{2}$,
we have
\begin{eqnarray}
\nonumber
\mathcal{L}V(i,\mathbf{x}) & \leq &
-2\alpha(i)\mathbf{x}^{T}\overline{L}^{2}\mathbf{x} +
\alpha^{2}(i)\left[c_{1}\lambda_{N}\left(\overline{L}\right)+
\frac{\lambda_{N}^{3}\left(\overline{L}\right)}{\lambda_{2}\left(\overline{L}\right)}
\mathbf{x}^{T}\overline{L}\mathbf{x}\right.\\
\nonumber
&&\hspace{5cm}\left.
+\frac{4N^{2}\lambda_{N}\left(\overline{L}\right)}{\lambda_{2}\left(\overline{L}\right)}
\mathbf{x}^{T}\overline{L}\mathbf{x}+
\frac{c_{2}\lambda_{N}\left(\overline{L}\right)}{\lambda_{2}\left(\overline{L}\right)}
\mathbf{x}^{T}\overline{L}\mathbf{x}\right]
\nonumber\\
\nonumber
& \leq &
-\alpha(i)\varphi\left(i,\mathbf{x}\right)+g(i)\left[1+V\left(i,\mathbf{x}\right)\right]
\end{eqnarray}
where
$\varphi(i,\mathbf{x})=2\mathbf{x}^{T}\overline{L}^{2}\mathbf{x},\:\:
g(i)=\alpha^{2}(i)\max\left(c_{1}\lambda_{N}(\overline{L}),\frac{\lambda_{N}^{3}(\overline{L})}{\lambda_{2}\left(\overline{L}\right)}
+\frac{4N^{2}\lambda_{N}(\overline{L})}{\lambda_{2}\left(\overline{L}\right)}
+\frac{c_{2}\lambda_{N}(\overline{L})}{\lambda_{2}}\right)
$.
It can be verified that $\varphi(i,\mathbf{x})$ and $g(i)$  satisfy the conditions for Theorem~\ref{MarkovProc}. Hence~(\ref{App0}).
\end{proof}
\begin{theorem}
\label{ConvProofGen2}Consider the $\mathcal{A-ND}$  algorithm under the
Assumptions~\textbf{1.2)}, \textbf{2.1)}, and \textbf{3)}. Then, there exists a.s.~a finite real random variable~$\mathbf{\theta}$ such that
\begin{equation}
\label{App6}
\mathbb{P}\left[\lim_{i\rightarrow\infty}\mathbf{x}(i)=\mathbf{\theta}\mathbf{1}\right]
= 1
\end{equation}
\end{theorem}
\begin{proof}
Note that $\mathbf{1}^TL\left(i,\mathbf{x}(i)\right)=\mathbf{0},\:\forall i$.
The proof then follows from Theorem~\ref{ConvProof1},
since the noise assumptions are the same.
\end{proof}

\begin{lemma}
\label{stat-thetaGen} Let $\theta$ be as given in
Theorem~\ref{ConvProofGen2} and $r$, the initial average, as given
in eqn.~(\ref{def_r}), under the Assumptions~\textbf{1.2)}, \textbf{2.1)},
and \textbf{3)}. Let the m.s.e.~$\zeta = \mathbb{E}\left[\mathbf{\theta} -
r\right]^{2}$.
 Then we have:
%
\begin{eqnarray}
\label{unb-theta_app}
\hspace{-2in}\mbox{\textbf{1)~Unbiasedness}:}\hspace{2in}\mathbb{E}\left[\theta\right]&=&r\\
%
\label{msbound_app}
\hspace{-2in}\mbox{\textbf{2)~M.S.E.~Bound}:}\hspace{2.135in}\zeta
&\leq& \frac{\eta}{N^{2}}\sum_{i\geq 0}\alpha^{2}(i)
\end{eqnarray}
%
\end{lemma}
\begin{proof}
Follows from Theorem~\ref{ConvProofGen2} and
Lemma~\ref{stat-theta}.
\end{proof}
It is possible to have results similar to
Theorem~\ref{ConvProofGen2} and Lemma~\ref{stat-thetaGen} under
 Assumption~\textbf{2.2)} on the noise. In
that case, we need exact mixing conditions on the sequence. Also,
Assumption~\textbf{2.2)} places no restriction on the growth rate of the variance of the noise component in the consensus
subspace. By Theorem~\ref{ConvProofgen}, we still get
a.s.~consensus, but the m.s.e.~may become unbounded, if no growth
restrictions are imposed.

\bibliographystyle{IEEEtran}
\bibliography{IEEEabrv,BibOptEqWeights}

\end{document}